\documentclass[final,1p,times,twocolumn]{elsarticle}

\usepackage{latexsym,,amssymb,amsmath,graphicx,epsf,bbm,float,subfig}
\usepackage{ifpdf}
\usepackage{epstopdf}
\usepackage{color}
\usepackage{tabularx}
\usepackage{stackrel}
\usepackage{stfloats}
\usepackage[percent]{overpic}
\usepackage{lineno,hyperref}
\usepackage{tabularx}
\usepackage{tikz}
\modulolinenumbers[5]

\journal{Journal of \LaTeX\ Templates}

\DeclareMathAlphabet\mathbfcal{OMS}{cmsy}{b}{n}

\newcommand{\defi}{:=}
\newcommand{\nn}{\nonumber}
\renewcommand{\vec}[1]{\mbox{$\mathbf{#1}$}}

\newcommand{\mSig}{\Sigma}
\newcommand{\mhSig}{\hat{\Sigma}}
\newcommand{\mbSig}{\bar{\Sigma}}

\newcommand{\vy}{\vec{y}}

\newcommand{\mH}{H}
\newcommand{\mhH}{\hat{H}}
\newcommand{\mA}{A}
\newcommand{\mB}{B}
\newcommand{\mP}{P}
\newcommand{\mbH}{\bar{H}}
\newcommand{\mbG}{\bar{G}}
\newcommand{\mtH}{\tilde{H}}
\newcommand{\mtG}{\tilde{G}}
\newcommand{\mT}{T}
\newcommand{\mG}{G}
\newcommand{\mK}{K}
\newcommand{\mC}{C}

\newcommand{\mD}{D}
\newcommand{\mM}{M}
\newcommand{\mN}{N}
\newcommand{\mL}{L}

\newcommand{\eye}{I}
\newcommand{\mZero}{\vec{0}}

\newcommand{\col}{\mathrm{col}}
\newcommand{\diag}{\mathrm{diag}}
\newcommand{\bx}{\pmb{x}}
\newcommand{\bn}{\pmb{n}}
\newcommand{\bu}{\pmb{u}}
\newcommand{\by}{\pmb{y}}
\newcommand{\br}{\pmb{r}}
\newcommand{\bm}{\pmb{m}}
\newcommand{\bdel}{\pmb{\delta}}

\newcommand{\cN}{{\cal N}}
\newcommand{\cC}{{\cal C}}
\newcommand{\vones}{\vec{1}}
\newcommand{\vzeros}{\vec{0}}

\newenvironment{psmallmatrix}
  {\left[\begin{smallmatrix}}
  {\end{smallmatrix}\right]}

\begin{document}

\begin{frontmatter}

\title{Team-Optimal Distributed MMSE Estimation in General and Tree Networks}

\author[uiuc]{Muhammed O. Sayin\corref{mycorrespondingauthor}}
\cortext[mycorrespondingauthor]{Corresponding author}
\ead{sayin2@illinois.edu}

\author[bilkent]{Suleyman S. Kozat}
\ead{kozat@ee.bilkent.edu.tr}

\author[uiuc]{Tamer Ba\c{s}ar}
\ead{basar1@illinois.edu}

\address[uiuc]{The Department of Electrical and Computer Engineering, University of Illinois at Urbana-Champaign, Champaign, IL 61801 USA}
\address[bilkent]{The Department of Electrical and Electronics Engineering, Bilkent University, Bilkent, Ankara 06800 Turkey}

\begin{abstract}
We construct team-optimal estimation algorithms over distributed networks for state estimation in the finite-horizon mean-square error (MSE) sense. Here, we have a distributed collection of agents with processing and cooperation capabilities. These agents observe noisy samples of a desired state through a linear model and seek to learn this state by interacting with each other. Although this problem has attracted significant attention and been studied extensively in fields including machine learning and signal processing, all the well-known strategies do not achieve team-optimal learning performance in the finite-horizon MSE sense. To this end, we formulate the finite-horizon distributed minimum MSE (MMSE) when there is no restriction on the size of the disclosed information, i.e., oracle performance, over an arbitrary network topology. Subsequently, we show that exchange of local estimates is sufficient to achieve the oracle performance only over certain network topologies. By inspecting these network structures, we propose recursive algorithms achieving the oracle performance through the disclosure of local estimates. For practical implementations we also provide approaches to reduce the complexity of the algorithms through the time-windowing of the observations. Finally, in the numerical examples, we demonstrate the superior performance of the introduced algorithms in the finite-horizon MSE sense due to optimal estimation.
\end{abstract}


\end{frontmatter}


\section{Introduction}
Over a distributed network of agents with measurement, processing and communication capabilities, we can have enhanced processing performance, e.g., fast response time, relative to the centralized networks by distributing the processing power over the networks \cite{banavar15, sayedMag,karMag, bes15}. Mainly, distributed agents observe the true state of the system through noisy measurements from different viewpoints, process the observation data in order to estimate the state, and communicate with each other to alleviate the estimation process in a fully distributed manner. Notably, the agents can respond to streaming data in an online manner by disclosing information among each other at certain instances. This framework is conveniently used to model highly complex structures from defense applications to social and economical networks \cite{krish11,ali2012, daron2011, jackson_book}. As an example, say that we have radar systems distributed over an area and seeking to locate hostile missiles, i.e., the location of the missile is the underlying state of the system. In that respect, distributed processing approach has vital importance in terms of detecting the missiles and reacting as fast as possible. In particular, even if the viewpoints of a few radar systems are blocked due to environmental obstacles, through the communication among the radar systems, each system should still be able to locate the missiles. Additionally, since each radar system not only collects measurements but also process them to locate the missiles, the overall system can respond to the missiles faster than a centralized approach in which measurements of all the radar systems are collected at a centralized unit and processed together.

Although there is an extensive literature on this topic, e.g., \cite{sayedMag, ali2012, daron2011, jackson_book, lopes2008, cattivelli2010, shahNips} and references therein, we still have significant and yet unexplored problems for disclosure and utilization of information among agents. Prior work has focused on the computationally simple algorithms that aim to minimize certain cost functions through the exchange of local estimates, e.g., diffusion or consensus based estimation algorithms \cite{sayedMag, lopes2008, karMag, giannakis09,sayin2013,sayin2014}, due to processing power related practical concerns. However, there is a trade-off in terms of computational complexity and estimation performance.

Formulating the optimal distributed estimation algorithms with respect to certain performance criteria is a significant and unexplored challenge. To this end, we consider here the distributed estimation problem as a team problem for distributed agents in which agents take actions, e.g., which information to disclose and how to construct the local estimate. This differs from the existing approaches in which agents exchange their local estimates. Furthermore, we address the optimality of exchanging local estimates with respect to the team problem over arbitrary network structures.

We examine the optimal usage of the exchanged information based on its content rather than a blind approach in which exchanged information is handled irrespective of the content as in the diffusion or consensus based approaches. In such approaches, the agents utilize the exchanged information generally through certain static combination rules, e.g., the uniform rule \cite{uniform}, the Laplacian rule \cite{scherber2004} or the Metropolis rule \cite{metropolis}. However, if the statistical profile of the measurement data varies over the network, i.e., each agent observes diverse signal-to-noise ratios, by ignoring the variation in noise, these rules yield severe degradation in the estimation performance \cite{sayedMag}. In such cases the agents can perform better even without cooperation \cite{sayedMag}. Therefore, the optimal usage of the exchanged information plays an essential role in performance improvement in the team problem.

Consider distributed networks of agents that observe noisy samples of an underlying state (possibly multi-dimensional) over a finite horizon. The agents can exchange information with only certain other agents at each time instant. In particular, agents cooperate with each other as a team according to a certain team cost depending on the agents' actions. To this end, each agent constructs a local estimate of the underlying state and constructs messages to disclose to the neighboring agents at each time instant. We particularly consider a quadratic cost function and that the underlying state and measurement noises are jointly Gaussian.

We note that restrictions on the sent messages, e.g., on the size of the disclosed information, has significant impact on the optimal team actions. We introduce the concept of the oracle performance, in which there is no restriction on the disclosed information. In that case, a time-stamped information disclosure can be team-optimal and we introduce the optimal distributed online learning (ODOL) algorithm using the time-stamped information disclosure. Through a counter example, we show that the oracle performance cannot be achieved through the exchange of local estimates. Then, we analytically show that over certain networks, e.g., tree networks, agents can achieve the oracle performance through the exchange of local estimates. We propose the optimal and efficient distributed online learning (OEDOL) algorithm, which is practical for real life applications and achieves the oracle performance over tree networks through the exchange of local estimates. Finally, we introduce the time windowing of the measurements in the team cost and propose a recursive algorithm, sub-optimal distributed online learning (SDOL) algorithm, combining the received messages linearly through time-invariant combination weights.

We can list our main contributions as follows: 1) We introduce a team-problem to minimize finite horizon mean square error cost function in a distributed manner. 2) We derive the ODOL algorithm achieving the oracle performance over arbitrary networks through time-stamped information exchange. 3) We address whether agents can achieve the oracle performance through the disclosure of local estimates. 4) We propose a recursive algorithm, the OEDOL algorithm, achieving the oracle performance over certain network topologies with tremendously reduced communication load. 5) We also formulate sub-optimal versions of the algorithms with reduced complexity. 6) We provide numerical examples demonstrating the significant gains due to the introduced algorithms.

The remainder of the paper is organized as follows. We introduce the team problem for distributed-MMSE estimation in Section 2. We study the tree networks, exploit the network topology to formulate the OEDOL algorithm that reduces the communication load and introduce cell structures, which is relatively more connected than tree networks, in Section 3. We propose the sub-optimal versions of the ODOL algorithm for practical implementations in Section 4. In Section 5, we provide numerical examples demonstrating significant gains due to the introduced algorithms. We conclude the paper in Section 6 with several remarks.

{\em Notation:} We work with real data for notational simplicity. ${\mathbb N}(0,.)$ denotes the multivariate Gaussian distribution with zero mean and designated covariance. For a vector $a$ (or matrix $A$), $a'$ (or $A'$) is its ordinary transpose. We denote the vector whose terms are all $1$s (or all $0$s) by $\vones$ (and $\vzeros$). We denote random variables by bold lower case letters, e.g., $\bx$. The operator $\mathrm{col}\{\cdot\}$ produces a column vector or a matrix in which the arguments of $\mathrm{col}\{\cdot\}$ are stacked one under the other. For a matrix $A$, $\mathrm{diag}\{A\}$ operator constructs a diagonal matrix with the diagonal entries of $A$. For a given set $\cN$, $\mathrm{diag}\{\cN\}$ creates a diagonal matrix whose diagonal block entries are elements of the set. The operator $\otimes$ denotes the Kronecker product.

\begin{figure}
  \centering
  \subfloat[First-order neighborhood]{\includegraphics[width=0.4\textwidth]{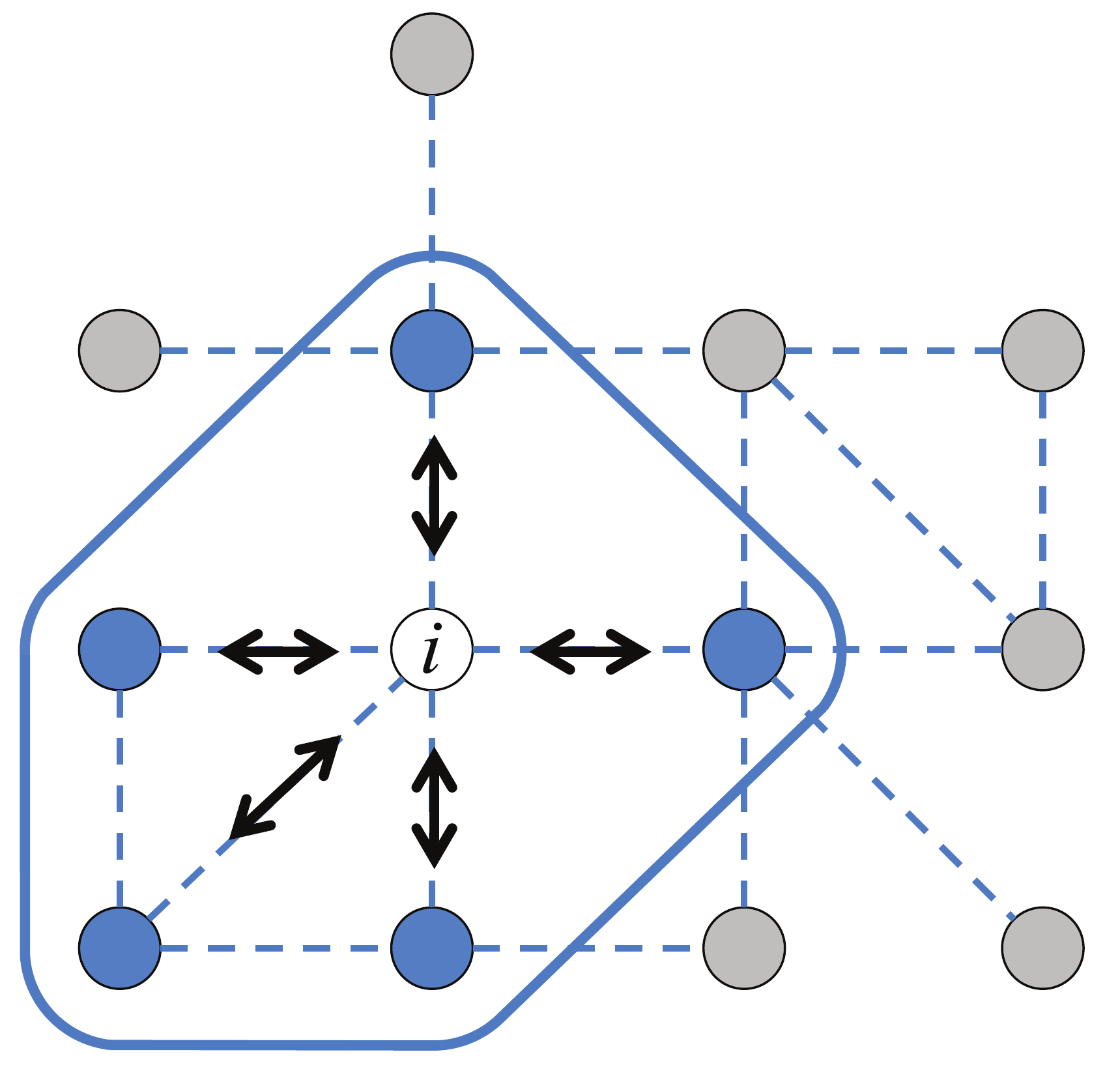}\label{fig:1a}}
  \hfil
  \subfloat[All neighborhoods]{\includegraphics[width=0.5\textwidth]{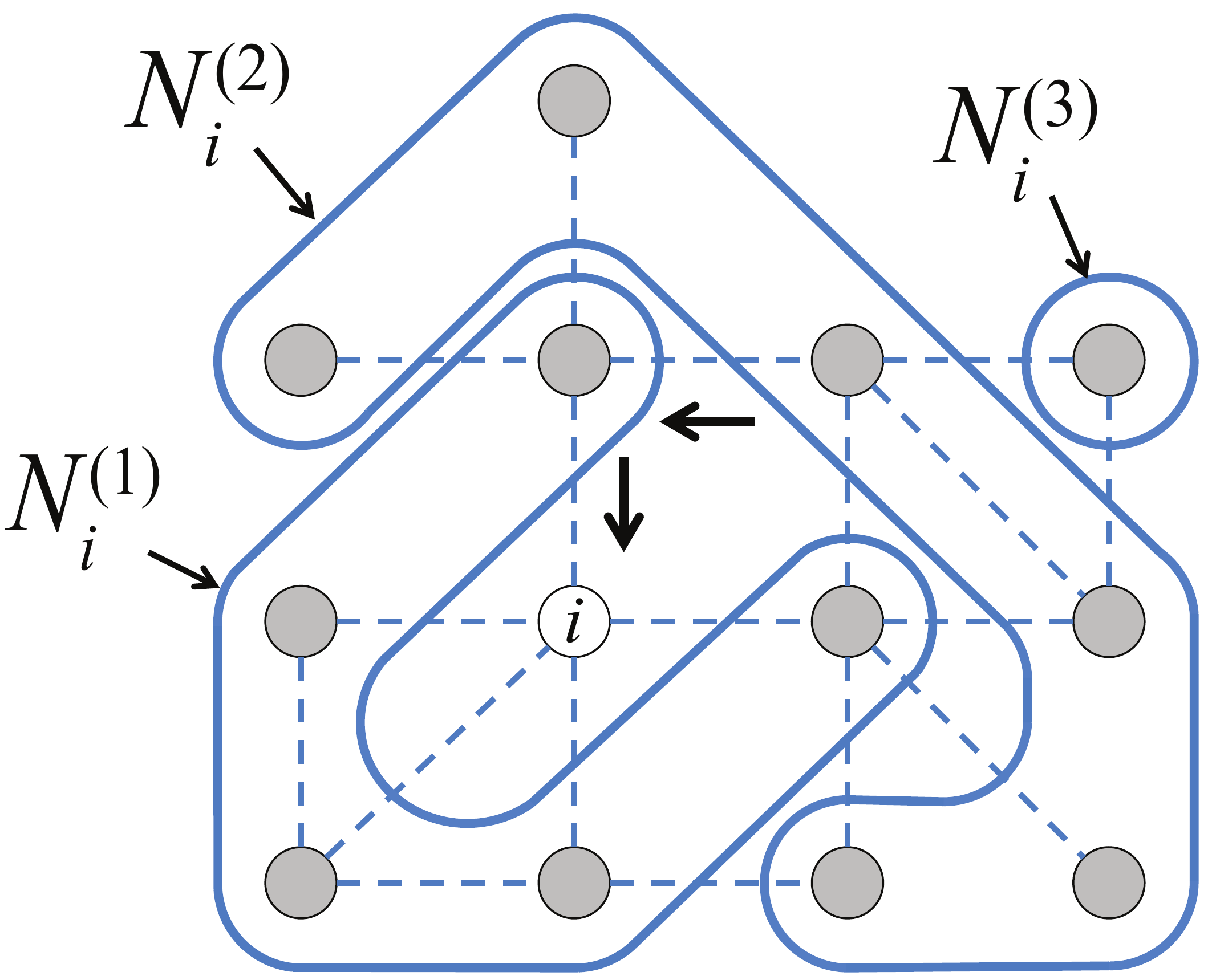}\label{fig:1b}}
  \caption{The neighborhoods of $i$th agent over the distributed network.}
  \label{fig:1}
\end{figure}

\section{Team Problem for Distributed-MMSE Estimation}
Consider a distributed network of $m$ agents with processing and communication capabilities. In Fig. \ref{fig:1}, we illustrate this network through an undirected graph, where the vertices and the edges correspond to the agents and the communication links across the network, respectively. For each agent $i$, we denote the set of agents whose information could be received at least after $k$ hops, i.e., $k$-hop neighbors, by $\cN_i^{(k)}$, and $\pi_i^{(k)} \defi \left|\cN_{i}^{(k)}\right|$ is the cardinality of $\cN_{i}^{(k)}$ (See Fig. \ref{fig:1b})\footnote{For notational simplicity, we define $\cN_i := \cN_i^{(1)}$ and $\pi_i := \pi_i^{(1)}$.}. We assume that  $\cN_i^{(0)} = \{i\}$ and $\cN_i^{(k)} = \varnothing$ for $k<0$. Note that the sequence of the sets $\{\cN_i^{(0)},\cN_i^{(1)},\cdots\}$ is a non-decreasing sequence such that $\cN_i^{(k)}\subseteq\cN_i^{(l)}$ if $k < l$.

Here, at certain time instants, the agents observe a noisy version of a time-invariant and unknown state vector $x \in \mathbbm{R}^p$ which is a realization of a Gaussian random variable $\bx$ with mean $\bar{x}$ and auto-covariance matrix $\mSig_x$. In particular, at time instant $t$, each agent $i$ observes a noisy version of the state as follows:
\[
y_{i,t} = H_i x + n_{i,t},\; t=1,\cdots,T,
\]
where $H_i \in \mathbbm{R}^{q\times p}$ is a matrix commonly known by all agents, and $n_{i,t} \in \mathbbm{R}^q$ is a realization of a zero-mean white Gaussian vector process $\{\bn_{i,t}\}$ with auto-covariance $\mSig_{n_i}$. Correspondingly, the observation $y_{i,t} \in \mathbbm{R}^q$ is a realization of the random process $\{\by_{i,t}\}$, where $\by_{i,t} = H_i\bx + \bn_{i,t}$ almost everywhere (a.e.). The noise $\bn_{i,t}$ is also independent of the state $\bx$ and the other noise parameters $\bn_{j,\tau}$, $j\neq i$ and $\tau \leq t$. We assume that the statistical profiles of the noise processes are common knowledge of the agents since they can readily be estimated from the data \cite{sayed_book}.

The agents have communication capabilities and at certain time instants, i.e., after each measurement, they can exchange information with the neighboring agents as seen in Fig. \ref{fig:1a}. Let $z_{i,j,t} \in \mathbb{R}^r$ denote the information disclosed by $i$ to $j$ at time $t$, and $r\geq 0$ is the size of the disclosed information. We assume that there exists a perfect channel between the agents such that the disclosed information can be transmitted with infinite precision. Therefore, we denote the information available to agent-$i$ at time $t$ by
\[
\delta_{i,t} = \Big\{y_{i,t},y_{i,\tau},z_{j,i,\tau},\mbox{ for } j \in \cN_i,\tau = 1,\cdots,t-1\Big\}
\]
and let $\sigma_{i,t}$ denote the sigma-algebra generated by the information set $\delta_{i,t}$. Furthermore, we define the set of all $\sigma_{i,t}$-measurable functions from $\mathbb{R}^{qt}\times\mathbb{R}^{r\pi_i(t-1)}$ to $\mathbb{R}^{r}$ by $\Gamma_{i,t}$. Importantly, here, which information to disclose is not determined a priori in the problem formulation. Let $\gamma_{i,j,t}$ be the decision strategy for $z_{i,j,t}$, then agent-$i$ chooses $\gamma_{i,j,t}$, $j \in \cN_i$, from the set $\Gamma_{i,t}$, i.e., $\gamma_{i,j,t} \in \Gamma_{i,t}$ and $\gamma_{i,j,t}(\delta_{i,t}) = z_{i,j,t}$, based on his/her objective.

In addition to the disclosed information $z_{i,j,t}$, $j\in\cN_i$, agent-$i$ takes action $u_{i,t} \in \mathbb{R}^q$, where the corresponding decision strategy $\eta_{i,t}$ is chosen from the set $\Omega_{i,t}$, which is the set of all $\sigma_{i,t}$-measurable functions from $\mathbb{R}^{qt}\times\mathbb{R}^{r\pi_i(t-1)}$ to $\mathbb{R}^{p}$, i.e., $\eta_{i,t}\in \Omega_{i,t}$ and $\eta_{i,t}(\delta_{i,t}) = u_{i,t}$. Here, we consider that the agents have a common cost function:
\[
\sum_{t=1}^T \sum_{j=1}^m \|x - u_{j,t}\|^2,
\]
where all actions $u_{i,t}$, $i=1,\cdots,m$ and $t=1,\cdots,T$ are costly, and agent-$i$ should take actions $u_{i,t}$ and $z_{i,j,t}$, $j=1,\cdots,\pi_i$ and $t=1,\cdots,T$, accordingly. Therefore, this corresponds to a team-problem, in which agent-$i$ faces the following minimization problem:
\begin{equation}
\min\limits_{\substack{\gamma_{i,j,t}\in\Gamma_{i,t},\eta_{i,t}\in\Omega_{i,t},\\ j\in\cN_i, t = 1,\cdots,T}} \sum_{t=1}^T \sum_{j=1}^m E\|\bx - \eta_{j,t}(\delta_{j,t})\|^2. \label{eq:cost1}
\end{equation}
We point out that both $\Gamma_{i,t}$ and $\Omega_{i,t}$ are infinite dimensional, i.e., \eqref{eq:cost1} is a functional optimization problem and the optimal strategies can be a nonlinear function of the available information. Furthermore, the agents should also construct the disclosed information accordingly since other agents' decisions $u_{j,t}$ directly depend on the disclosed information.

\subsection{A Lower Bound}

In order to compute the team optimal strategies, we first construct a lower bound on the performance of the agents by removing the limitation on the size of the disclosed information, i.e., $r\rightarrow \infty$. In that case, the following proposition provides an optimal information disclosure strategy.

\begin{figure}[t!]
  \centering
  \includegraphics[width=0.8\textwidth]{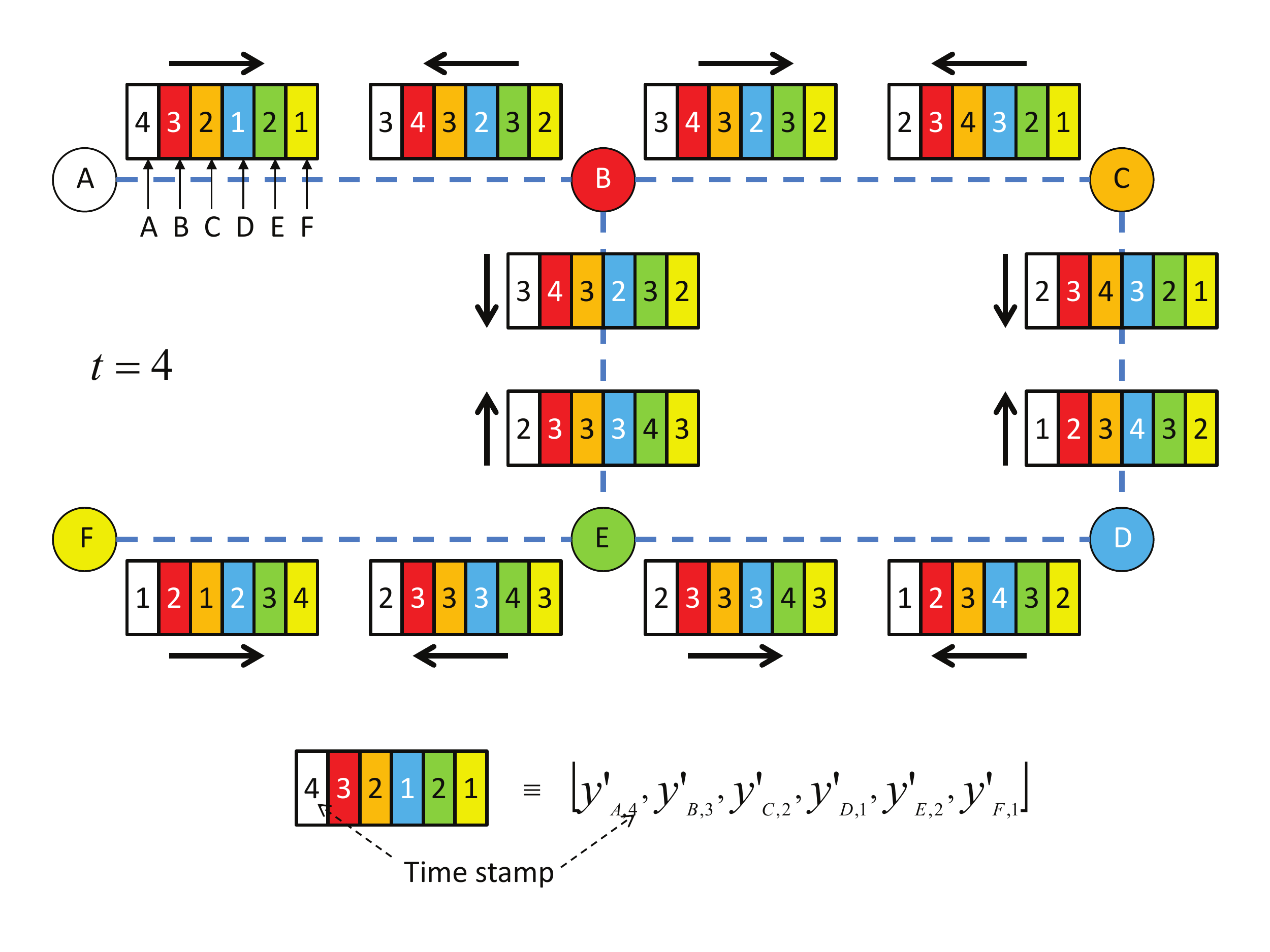}\\
  \caption{Time stamped information disclosure over a network of $6$ agents at time instant $t=4$.}\label{fig:exa}
\end{figure}

{\bf Proposition 2.1.}
{\em When $r\rightarrow \infty$, a time stamped information disclosure strategy, in which agents transmit the most current version of the available information (e.g., see Fig. \ref{fig:exa}), can lead to the team-optimal solution.}

{\em Proof.} Through the time stamped information disclosure, each agent can obtain the measurements of the other agents separately in a connected network. However, the measurements of the non-neighboring agents could only be received after certain hops due to the partially connected structure, i.e., certain agents are not directly connected. As an example, the disclosed information of $j \in \cN_i^{(2)}$ reaches to $i$ by passing through two communication links as seen in Fig. \ref{fig:1b}. In particular, this case assumes that each agent has access to full information from the other agents, albeit with certain hops, and corresponds to the direct aggregation of all measurements across the network at each agent.

Correspondingly, at time $t$, all the information aggregated at $i$th agent is given by
\begin{equation}\label{eq:deltao}
\delta_{i,t}^o :=
\left\{\left\{y_{i,\tau}\right\}^{\tau \leq t},\left\{y_{j,\tau}\right\}^{\tau \leq t-1}_{j\in \cN_i^{(1)}},\cdots,\left\{y_{j,\tau}\right\}^{\tau \leq t-\kappa_i}_{j\in \cN_{i}^{(\kappa_i)}}\right\},
\end{equation}
where the information from the furthest agent is received at least after $\kappa_i$ hops. Therefore, a time-stamped information disclosure strategy can lead to the team-optimal solution. \hfill $\square$

Let $\sigma_{i,t}^o$ denote the sigma-algebra generated by the information set $\delta_{i,t}^o$ and $\Omega_{i,t}^o$ be the set of all $\sigma_{i,t}^o$-measurable functions from $\mathbb{R}^{qt}\times\mathbb{R}^{q\pi_i(t-1)}\times\cdots\times\mathbb{R}^{q\pi_i^{\kappa_i}|t-\kappa_i|_+}$ (where $|t-\kappa_i|_+ = 0$ if $t-\kappa_i < 0$) to $\mathbb{R}^p$. Then, agent-$i$ faces the following minimization problem:
\begin{equation}
\min\limits_{\substack{\eta_{i,t}\in\Omega^o_{i,t},\\ t = 1,\cdots,T}} \sum_{t=1}^T \sum_{j=1}^m E\|\bx - \eta_{j,t}(\delta_{j,t}^o)\|^2, \label{eq:costlow}
\end{equation}
which is equivalent to
\begin{equation}
\sum_{t=1}^T\min\limits_{\substack{\eta_{i,t}\in\Omega^o_{i,t}}} E\|\bx - \eta_{i,t}(\delta_{i,t}^o)\|^2\label{eq:costlow2}
\end{equation}
since $\gamma_{i,j,t}$, $j\in \cN_i$, is set to the time-stamped strategy and $\eta_{i,t}$ has impact only on the term $E\|\bx - \eta_{i,t}(\delta_{i,t}^o)\|^2$. Let $\bdel_{i,t}^o$ be defined by
\[
\bdel_{i,t}^o :=
\left\{\left\{\by_{i,\tau}\right\}^{\tau \leq t},\left\{\by_{j,\tau}\right\}^{\tau \leq t-1}_{j\in \cN_i^{(1)}},\cdots,\left\{\by_{j,\tau}\right\}^{\tau \leq t-\kappa_i}_{j\in \cN_{i}^{(\kappa_i)}}\right\}
\]
Then, team optimal strategy in the lower bound, i.e., oracle strategy, $\eta_{i,t}^o$ and the corresponding action $u_{i,t}^o$ are given by
\begin{align}
\eta_{i,t}^o(\delta_{i,t}^o) = u_{i,t}^o = E[\bx|\bdel_{i,t}^o = \delta_{i,t}^o]\label{eq:x}
\end{align}
and we define $\bu_{i,t}^o := E[\bx|\bdel_{i,t}^o]$.

\subsection{ODOL Algorithm}
Since the state and the observation noise are jointly Gaussian random parameters, we can compute \eqref{eq:x} through a Kalman-like recursion \cite{moore_book}. Therefore, we provide the following ODOL algorithm. We introduce a difference set $\Delta_{i,t} := \delta_{i,t}^o \backslash \delta_{i,t-1}^o = \big\{y_{i,t},\left\{y_{j,t-1}\right\}_{j\in \cN_i^{(1)}},\cdots,\left\{y_{j,t-\kappa_i}\right\}_{j\in \cN_{i}^{(\kappa_i)}}\big\}$ and a vector $w_{i,t} = \col\{\Delta_{i,t}\}$. Then, for $t\geq 1$ the iterations of the ODOL algorithm are given by
\begin{align*}
&u_{i,t}^o = \left(\eye - \mK_{i,t}\mbH_{i}\right)u_{i,t-1}^o + \mK_{i,t}w_{i,t}, \\
&\mK_{i,t} = \mhSig_{i,t-1}\mbH_{i}^T\left(\mbH_{i}\mhSig_{i,t-1}\mbH_{i}^T + \mbSig_{n_i}\right)^{-1},\\
&\mhSig_{i,t} = \left(\eye - \mK_{i,t}\mbH_{i}\right)\mhSig_{i,t-1},
\end{align*}
where\footnote{If the inverse fails to exist, a pseudo inverse can replace the inverse \cite{moore_book}.} $u_{i,0}^o = \bar{x}$, $\mhSig_{i,0} = \mSig_x$, $\mbH_i \defi \left(\mP_i \otimes \eye_q \right) \mH$, $\mH \defi \mathrm{col}\left\{\mH_1,\cdots,\mH_m\right\}$, $\mbSig_{n_i} \defi \left(\mP_i \otimes \eye_p\right)\mSig_n \left(\mP_i \otimes \eye_p\right)^T$, $\mSig_n \defi \mathrm{diag} \left\{\mSig_{n_1},\cdots,\mSig_{n_m}\right\}$, and $\mP_i$ is the corresponding permutation matrix.

We point out that this is a lower bound on the original cost function \eqref{eq:cost1}, i.e.,
\begin{equation}
\sum_{t=1}^T \sum_{i=1}^m  E\|\bx - u_{i,t}^o\|^2 \leq \min\limits_{\substack{\gamma_{i,j,t}\in\Gamma_{i,t},\eta_{i,t}\in\Omega_{i,t},\\ i=1,\cdots,m;j\in\cN_i; t = 1,\cdots,T}} \sum_{t=1}^T \sum_{j=1}^m E\|\bx - \eta_{j,t}(\delta_{j,t})\|^2,\label{eq:lower}
\end{equation}
where we substitute team-optimal action (when $r\rightarrow \infty$) $u_{i,t}^o$ back into \eqref{eq:costlow2} and sum over $t=1,\cdots,T$ and $i=1,\cdots,m$. However, the lower bound is not necessarily tight depending on $r$. By Proposition 2.1, time-stamped information disclosure strategy, in which the size of the disclosed information is $q\times m$, yields the oracle solution. This implies that when $r \geq qm$, the lower bound is tight. Furthermore, team optimal solutions are linear in the available information and can be constructed through the recursive algorithm ODOL. However, $qm$ is linear in the number of agents, $m$, and in large networks this can cause excessive communication load yet communication load is crucial for the applicability of the distributed learning algorithms \cite{sayin2014,sayin2013}. Therefore, in the following section, we provide a sufficient condition on the size of the disclosed information, which depends on the network structure (rather than its size), in order to achieve the lower bound \eqref{eq:lower}.

\section{Distributed-MMSE Estimation with Disclosure of Local Estimate}
In the conventional distributed estimation algorithms, e.g., consensus and diffusion approaches, agents disclose their local estimates, which have size $p$ (note that this does not depend on the network size). The following example addresses whether the disclosure of local estimates can achieve the lower bound \eqref{eq:lower} or not.

\begin{figure}[t!]
  \centering
  \includegraphics[width=1in]{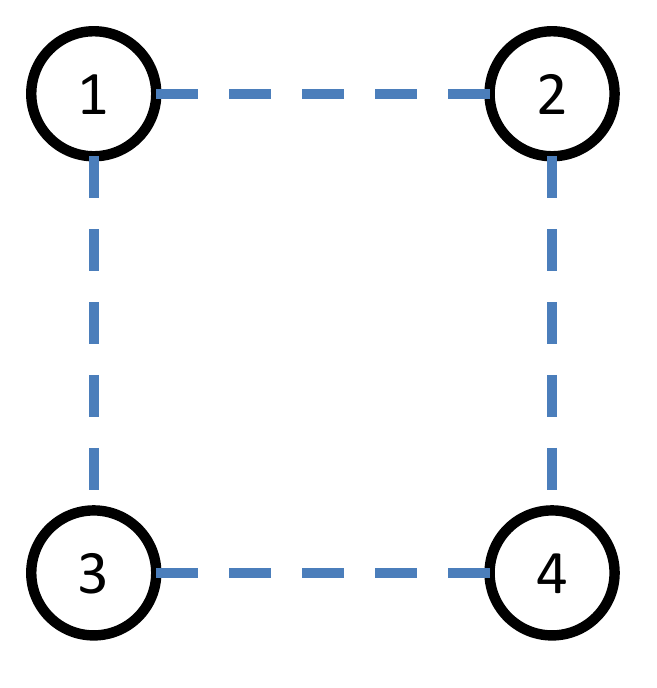}\\
  \caption{A cycle network of $4$ agents.}\label{fig:cycle}
\end{figure}

\subsection{A Counter Example}
Consider a cycle network of $4$ agents as seen in Fig. \ref{fig:cycle}, where $p=q=1$, $H_i = 1$, $\Sigma_{n_i} = \sigma_{n}^2$, for $i = 1,\cdots,4$, and $\Sigma_x = \sigma_x^2$. We aim to show that agent-$1$'s oracle action at time $t=3$, i.e., $u_{1,3}^o$, cannot be constructed through the exchange of local estimates.

At time $t = 2$, agent-$2$ and agent-$3$ have the following oracle actions:
\begin{align}
u_{2,2}^o &= E[\bx | \by_{2,2} = y_{2,2}, \by_{2,1} = y_{2,1}, \by_{1,1} = y_{1,1}, \by_{4,1} = y_{4,1}],\nn\\
&= \frac{\sigma_x^2}{4(\sigma_{x}^2+\sigma_n^2)}(y_{2,2}+y_{2,1}+y_{1,1}+y_{4,1}),\label{eq:est2}\\
u_{3,2}^o &= E[\bx | \by_{3,2} = y_{3,2}, \by_{3,1} = y_{3,1}, \by_{1,1} = y_{1,1}, \by_{4,1} = y_{4,1}].\nn\\
&= \frac{\sigma_x^2}{4(\sigma_{x}^2+\sigma_n^2)}(y_{3,2}+y_{3,1}+y_{1,1}+y_{4,1}).\label{eq:est3}
\end{align}
Note that since there are two hops between agents $2$ and $3$, at $t=2$, agents do not have access to each other's any measurement yet. At time $t=3$, agent-$1$'s oracle action is given by
\begin{align}
u_{1,3}^o = E[\bx | &\by_{1,3} = y_{1,3}, \by_{1,2} = y_{1,2}, \by_{1,1} = y_{1,1},\nn\\
&\by_{2,2} = y_{2,2}, \by_{2,1} = y_{2,1},\nn\\
&\by_{3,2} = y_{3,2}, \by_{3,1} = y_{3,1},\nn\\
&\by_{4,1} = y_{4,1}],\nn\\
= \frac{\sigma_x^2}{8(\sigma_{x}^2+\sigma_n^2)}&(y_{1,3}+y_{1,2}+y_{1,1}+y_{2,2}+y_{2,1}+y_{3,2}+y_{3,1}+y_{4,1}).\nn
\end{align}
Assume that $u_{1,3}^o$ can be obtained through the exchange of local estimates:
\begin{align}
\hat{u}_{1,3} := E[\bx | &\by_{1,3} = y_{1,3}, \by_{1,2} = y_{1,2}, \by_{1,1} = y_{1,1},\nn\\
 &\bu_{2,2}^o = u_{2,2}^o, \bu_{2,1}^o = u_{2,1}^o,\nn\\
 &\bu_{3,2}^o = u_{3,2}^o, \bu_{3,1}^o = u_{3,1}^o].\label{eq:new}
\end{align}
Since all parameters are jointly Gaussian, the local estimates are also jointly Gaussian, $\hat{u}_{1,3}$, is linear in $u_{2,2}^o$ and $u_{3,2}^o$. Furthermore, the measurements $y_{2,2}$, $y_{3,2}$, and $y_{4,1}$ are only included in $u_{2,2}^o$ and $u_{3,2}^o$. Therefore, we obtain
\begin{align}
\hat{u}_{1,3} &= \cdots + \alpha u_{2,2}^o + \beta u_{3,2}^o\nn\\
&= \cdots + \alpha \frac{\sigma_x^2}{4(\sigma_{x}^2+\sigma_n^2)} (y_{2,2} + y_{4,1} + \cdots) + \beta \frac{\sigma_x^2}{4(\sigma_{x}^2+\sigma_n^2)} (y_{3,2} + y_{4,1} + \cdots),\nn
\end{align}
where $\cdots$ refers to the other terms. However, the equality of $\hat{u}_{1,3}$ and $u_{1,3}^o$ implies $\alpha = \beta = 1/2$ due to the combination weights of $y_{2,2}$ and $y_{3,2}$, respectively, and $\alpha + \beta = 1/2$ due to the combination weight of $y_{4,1}$, which leads to a contradiction. Hence, which information to disclose over arbitrary networks for team-optimal solutions should be considered elaborately. In the following, we analytically show that lower bound could be achieved through the disclosure of local estimates over ``tree networks".

\subsection{Tree Networks}
A network has a ``tree structure" if its corresponding graph is a {\em tree}, i.e., connected and undirected without any cycles \cite{skiena_book}. As an example, the conventional star or line networks have tree structures. We remark that for an arbitrary network topology we can also construct the spanning tree of the network and eliminate the cycles.
In the literature, there exists numerous distributed algorithms for minimum spanning tree construction \cite{wu,gallager1983,peleg_book,elkin2004,khan2009}.

Importantly, the following theorem shows that over tree networks we can achieve the performance of the oracle algorithm through the disclosure of the local estimates only.

{\bf Theorem 3.1:}
{\em Consider the team-problem over a tree network, in which $r = p$. Then, exchange of local estimates can lead to the team-optimal solution, i.e., agents can achieve the oracle performance.}

{\em Proof:}
Initially, agent-$i$ has access to $y_{i,1}$ only and the oracle action is $u_{i,1}^o = E[\bx | \by_{i,1} = y_{i,1}]$. At time $t = 2$, the oracle action is given by
\begin{equation}\label{eq:x1}
u_{i,2}^o = E\left[\bx | \left\{\by_{i,\tau} = y_{i,\tau}\right\}_{\tau=1,2},\left\{\by_{j,1}=y_{j,1}\right\}_{j\in \cN_i}\right],
\end{equation}
which can be written as
\begin{align}
u_{i,2}^o &= E\left[\bx | \left\{\by_{i,\tau}=y_{i,\tau}\right\}_{\tau=1,2},\left\{E[\bx | \by_{j,1}=y_{j,1}]\right\}_{j\in \cN_i}\right],\nn\\
         &= E\left[\bx | \left\{\by_{i,\tau}=y_{i,\tau}\right\}_{\tau=1,2},\left\{\bu_{j,1}^o = u_{j,1}^o\right\}_{j \in \cN_i} \right].\label{eq:x11}
\end{align}
This implies that for $t=1$ and $t=2$, the oracle performance can be achieved through the disclosure of local estimate. Therefore, we can consider the oracle action \eqref{eq:x1} even though agents disclose their local estimate instead of time-stamped information disclosure.

\begin{figure}
  \centering
  \includegraphics[width=3in]{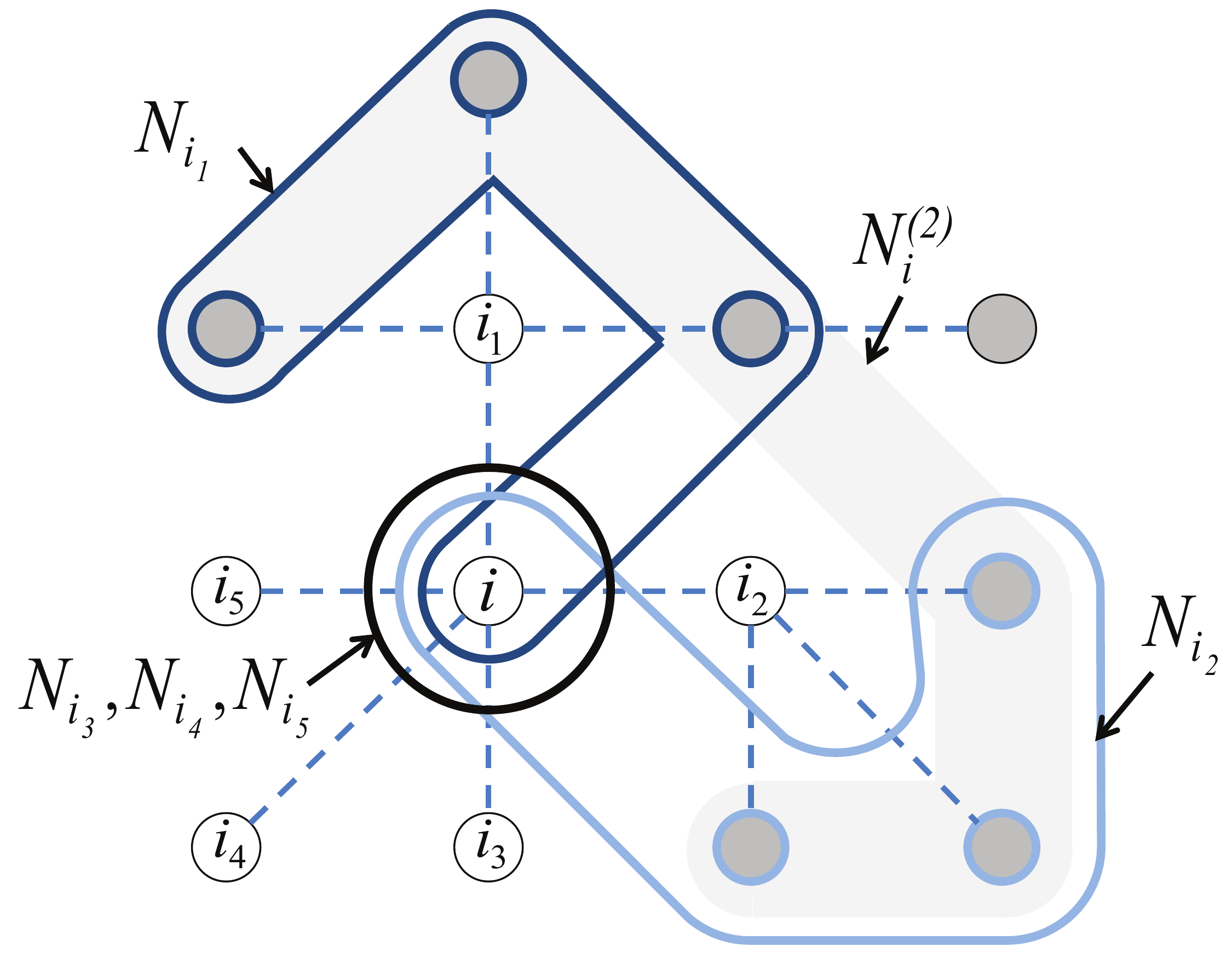}\\
  \caption{An example tree network. Notice the eliminated links from Fig. \ref{fig:1} to avoid multi path information diffusion.}\label{fig:tree}
\end{figure}

As seen in Fig. \ref{fig:tree}, over a tree network, for $k \in \{1,\cdots,\kappa_i\}$ we have
\begin{equation}
\cN_{i}^{(k)} = \bigcup_{j \in \cN_i} \left(\cN_{i}^{(k)}\cap \cN_{j}^{(k-1)}\right). \label{eq:cup}
\end{equation}
Note that the sets in \eqref{eq:cup} are disjoint as
\begin{equation}\label{eq:inter}
\left(\cN_{i}^{(k)} \cap \cN_{j_1}^{(k-1)}\right)\cap\left(\cN_{i}^{(k)} \cap \cN_{j_2}^{(k-1)}\right) = \varnothing
\end{equation}
for all $j_1,j_2 \in \cN_i$ and $j_1 \neq j_2$. Notably, over a tree network, by \eqref{eq:inter}, we can partition the collection set of the measurements received after at least $k$-hops as follows
\begin{align}
\left\{y_{j,\tau}\right\}_{j \in \cN_i^{(k)}} = \Bigg\{&\left\{y_{j,\tau}\right\}_{j \in \cN_i^{(k)}\cap \cN_{j_1}^{(k-1)}},\nn\\
&\cdots,\left\{y_{j,\tau}\right\}_{j \in \cN_i^{(k)}\cap \cN_{j_{\pi_i}}^{(k-1)}}\Bigg\}.\label{eq:1}
\end{align}

In the time-stamped information disclosure, at time $t=3$, agent-$i$ has access to $\delta_{i,3}^o$, defined in \eqref{eq:deltao}. We denote the set of new measurements received by $i$ over $j$ at time $t=2$ by
\begin{align}
\Delta_{j,i,2}^o \defi \Big\{\overbrace{\left\{y_{k,2}\right\}_{k \in \cN_i^{(1)}\cap \cN_j^{(0)}}}^{= y_{j,2}},\left\{y_{k,1}\right\}_{k \in \cN_i^{(2)}\cap \cN_j^{(1)}}\Big\},\nn
\end{align}
which can also be written as
\begin{equation}
\Delta_{j,i,2}^o = \delta_{j,2}^o \setminus \Big\{\overbrace{y_{j,1}}^{= \delta_{j,1}^o},\underbrace{y_{i,1}}_{=\Delta_{i,j,1}^o=\delta_{i,1}^o}\Big\}\label{eq:3},
\end{equation}
where we exclude the information sent by $i$ to $j$ at time $t=1$, i.e., $y_{i,1}$. Then, we can write the accessed information as the union of new measurement $y_{i,3}$, new measurements received over the neighboring agents and the accessed information at time $t=2$ as follows:
\begin{equation}\label{eq:2}
\delta_{i,3}^o = \left\{y_{i,3},\Delta_{j_1,i,2}^o,\cdots,\Delta_{j_{\pi_i},i,2}^o,\delta_{i,2}^o\right\}.
\end{equation}
Note that the sets on the right hand side of \eqref{eq:2} are disjoint due to tree structure. Furthermore, by \eqref{eq:3} and \eqref{eq:2}, the sigma-algebra generated by $\delta_{i,3}^o$ is equivalent to the sigma-algebra generated by the set $\left\{\{y_{i,\tau}\}^{\tau\leq 3}, \{\delta_{j,\tau}^o\}_{j\in\cN_i}^{\tau \leq 2}\right\}$. Since $u_{j,t}^o = E[\bx|\bdel_{j,t}^o = \delta_{j,t}^o]$, we obtain
\begin{align}
u_{i,3}^o = E\left[\bx | \left\{\by_{i,\tau}=y_{i,\tau}\right\}^{\tau \leq 3},\left\{\bu_{j,\tau}^o = u_{j,\tau}^o\right\}_{j \in \cN_i}^{\tau \leq 2}\right].\label{eq:x22}
\end{align}

By \eqref{eq:3}, we have
\begin{equation}\label{eq:4}
\Delta_{j,i,t}^o = \delta_{j,t}^o \setminus \Big\{\delta_{j,t-1}^o \cup \Delta_{i,j,t-1}^o\Big\},
\end{equation}
which implies that for $t \geq 2$, $\Delta_{j,i,t}^o$ is constructible from $\delta_{i,\tau}^o$ and $\delta_{j,\tau}^o$ for $\tau \leq t$. Hence, by induction, we conclude that the lower bound can be achieved through the exchange of local estimates.\hfill $\square$

{\bf Remark 3.1:} {\em When the expectation of the state is conditioned on infinite number of observations over even a constructed spanning tree, only a finite number of the observations is missing compared to the case over a fully connected network. Hence, even if we construct the spanning tree of that network, we would still achieve the lower bound over a fully connected (or centralized) network asymptotically. As an illustrative example, in Fig. \ref{fig:comp_terminal}, we observe that the MMSE performance over the fully connected, star and line networks are asymptotically the same. Similarly, in \cite{chenConf, chenI, chenII}, the authors show that the performance of the diffusion based algorithms could approach the performance of a fully connected network under certain regularity conditions.}

In the sequel, we propose the OEDOL algorithm that achieves the lower bound over tree networks iteratively.


\subsection{OEDOL Algorithm}

By Theorem 3.1, over a tree network, oracle action can be constructed by
\begin{align}
u_{i,t}^o = E\left[\bx | \left\{\by_{i,\tau}=y_{i,\tau}\right\}^{\tau \leq t},\left\{\bu_{j,\tau}^o = u_{j,\tau}^o\right\}_{j \in \cN_i}^{\tau \leq t-1}\right]\label{eq:full3}
\end{align}
through the disclosure of oracle actions, i.e., local estimates. We remark that $u_{i,t}^o$ is linear in the previous actions $u_{i,\tau}^o$, $\tau \leq t-1$. In order to extract new information, i.e., innovation part, we need to eliminate the previously received information at each instant on the neighboring agents. This brings in additional computational complexity. On the contrary, agents can just send the new information compared to the previously sent information, e.g., $s_{i,t}$. Note that here agents disclose the same information to the neighboring agents. Since we are conditioning on the linear combinations of the conditioned variables without effecting their spanned space, i.e., $s_{i,t}$ is computable from $u_{i,\tau}^o$ for $\tau \leq t$ and vice versa, agents can still achieve the oracle performance by reduced computational load, yet.

At time $t$, agent-$i$ receives local measurement $y_{i,t}$ and sent information from the neighboring agents, $r_{i,t} := \col\{s_{j_1,t-1},\cdots,s_{j_{\pi_i},t-1}\}$. We aim to determine the content of the received information $r_{i,t}$ to extract the innovation within them and utilize this innovation in the update of the oracle action.

Initially, at time $t=1$, agent-$i$ has only access to the local measurement $y_{i,1}$. Then, the oracle action is given by
\[
u_{i,1}^o = (\eye - \mSig_{x}H_i'(H_i\mSig_x H_i' + \mSig_{n_i})^{-1}H_i)\bar{x} + \mSig_{x}H_i'(H_i\mSig_xH_i' + \mSig_{n_i})^{-1} y_{i,1}.
\]
Let $u_{i,0}^o = \bar{x}$ and $\mhSig_{i,0} = \mSig_x$, and set $B_{i,1} = \mhSig_{i,0}H_i'(H_i\mhSig_{i,0}H_i' + \mSig_{n_i})^{-1}$ and $A_{i,1} = \eye - B_{i,1}H_i$. Then, we obtain
\begin{align}
&u_{i,1}^o = A_{i,1}u_{i,0}^o + B_{i,1}y_{i,1},\nn\\
&\mhSig_{i,1} = A_{i,1}\mhSig_{i,0}.\nn
\end{align}
Next, instead of sending $u_{i,1}^o$, agent-$i$ sends to the neighboring agents, $j\in\cN_i$,
\begin{align}
s_{i,1} &= u_{i,1}^o - A_{i,1}u_{i,0}^o\nn\\
&= B_{i,1}y_{i,1}.\nn
\end{align}
Correspondingly, at time $t=2$, agent-$i$ receives $y_{i,2}$ and $r_{i,2}$. Let $\br_{i,2}$ be the corresponding random vector. Then, conditioning the state and the received information on the previously available information $\by_{i,1} = y_{i,1}$, we have
\[
\begin{psmallmatrix} \bx \\ \by_{i,2} \\ \br_{i,2} \end{psmallmatrix}\Big| \by_{i,1}=y_{i,1} \sim \mathbb{N} \left(\begin{psmallmatrix}  u_{i,1}^o \\ H_iu_{i,1}^o \\ \mbH_{i,1}u_{i,1}^o \end{psmallmatrix}, \begin{psmallmatrix} \mhSig_{i,1} & \mhSig_{i,1}H_i' & \mhSig_{i,1}\mbH_{i,1}' \\ H_i\mhSig_{i,1} & H_i\mhSig_{i,1}H_i' + \mSig_{n_i} & H_i\mhSig_{i,1}\mbH_{i,1}' \\ \mbH_{i,1}\mhSig_{i,1} & \mbH_{i,1}\mhSig_{i,1}H_i' & \mbH_{i,1}\mhSig_{i,1}\mbH_{i,1}' + \mbG_{i,1} \end{psmallmatrix}\right),
\]
where $\mbH_{i,1} := \col\left\{B_{j_1,1}H_{j_1},\cdots,B_{j_{\pi_i},1}H_{j_{\pi_i}}\right\}$ and $\mbG_{i,1} = \diag\{G_{i,1}\}$, where $G_{i,1} := \col\left\{B_{j_1,1}\mSig_{n_{j_1}}B_{j_1,1}',\cdots, B_{j_{\pi_i},1} \mSig_{n_{j_{\pi_i}}}B_{j_{\pi_i},1}'\right\}$.

Let $\mtH_{i,1} := \col\{H_i,\mbH_{i,1}\}$ and $\mtG_{i,1} := \diag\{\Sigma_{n_i},\mbG_{i,1}\}$ and set
\begin{align}
\begin{bmatrix} B_{i,2} & C_{i,2} \end{bmatrix} &= \mhSig_{i,1}\mtH_{i,1}'\left(\mtH_{i,1}\mhSig_{i,1}\mtH_{i,1}' + \mtG_{i,1}\right)^{-1},\label{eq:bc}\\
A_{i,2} &= \eye - B_{i,2}\mH_i - C_{i,2}\mbH_{i,1}.\nn
\end{align}
Then, we obtain
\begin{align}
&u_{i,2}^o = A_{i,2} u_{i,1}^o + B_{i,2} y_{i,2} + C_{i,2} r_{i,2},\label{eq:ui2}\\
&\mhSig_{i,2} = A_{i,2}\mhSig_{i,1}\nn
\end{align}
and agent-$i$ sends
\begin{align}
s_{i,2} &= u_{i,2}^o - A_{i,2}u_{i,1}^o\nn\\
&= B_{i,2}y_{i,2} + \sum_{j\in\cN_i} C_{i,2}^{(j)} \underbrace{B_{j,1}y_{j,1}}_{=s_{j,1}},\nn
\end{align}
where $C_{i,2}^{(j)}$ denotes the corresponding $j$th block of $C_{i,2}$. Therefore, at time $t=3$, agent-$i$ receives from $j\in\cN_{i}$:
\begin{equation}
s_{j,2} = B_{j,2}y_{j,2} + \sum_{k\in\cN_j\setminus i} (C_{j,2}^{(k)}B_{j,1}y_{j,1}) + C_{j,2}^{(i)}B_{i,1}y_{i,1}.\label{eq:s2}
\end{equation}
Since the last term on the right hand side of \eqref{eq:s2} is known by $i$, we have
\begin{equation}
E[\br_{i,2}|\bdel_{i,2}^o = \delta_{i,2}^o] = \overbrace{\begin{psmallmatrix} B_{j_1,2} + \sum_{k\in\cN_{j_1}\setminus i} C_{j_1,2}^{(k)} B_{k,2} \\ \vdots \\ B_{j_{\pi_i},2} + \sum_{k\in\cN_{j_{\pi_i}}\setminus i} C_{j_{\pi_i},2}^{(k)} B_{k,2} \end{psmallmatrix}}^{=\mbH_{i,2}} u_{i,2}^o + D_{i,2}s_{i,1},\label{eq:r2}
\end{equation}
where $D_{i,2} := \col\{C_{j_1,2}^{(i)},\cdots, C_{j_{\pi_i},2}^{(i)}\}$, and
\begin{equation}
G_{i,2} = \begin{psmallmatrix} B_{j_1,2}\mSig_{n_{j_1}}B_{j_1,2}' + \sum_{k\in\cN_{j_1}\setminus i} C_{j_1,2}^{(k)}B_{k,1}\mSig_{n_{k}}B_{k,1}'(C_{j_1,2}^{(k)})' \\ \vdots \\ B_{j_{\pi_i},2}\mSig_{n_{j_{\pi_i}}}B_{j_{\pi_i},2}' + \sum_{k\in\cN_{j_{\pi_i}}\setminus i} C_{j_{\pi_i},2}^{(k)}B_{k,1}\mSig_{n_{k}}B_{k,1}'(C_{j_{\pi_i},2}^{(k)})' \end{psmallmatrix}\label{eq:G2}
\end{equation}
By \eqref{eq:ui2}, \eqref{eq:r2}, and \eqref{eq:G2}, the next oracle action $u_{i,3}^o$ is given by
\[
u_{i,3}^o = A_{i,3} u_{i,2}^o + B_{i,3}y_{i,3} + C_{i,3}(r_{i,3} - D_{i,2}s_{i,1}).
\]
Subsequently, agent-$i$ sends $s_{i,3} = u_{i,3}^o - A_{i,3}u_{i,2}^o$ and the received information from $j \in \cN_i$ yields
\begin{align}
s_{j,3} =& B_{j,3}y_{j,3} + \sum_{k\in\cN_j} C_{j,3}^{(k)}\left(B_{k,2}y_{k,2} + \sum_{l\in\cN_k\setminus j} C_{k,2}^{(l)}B_{l,1}y_{l,1}\right)\nn\\
=& B_{j,3}y_{j,3} + \sum_{k\in\cN_j\setminus i} C_{j,3}^{(k)}\left(B_{k,2}y_{k,2} + \sum_{l\in\cN_k\setminus j} C_{k,2}^{(l)}B_{l,1}y_{l,1}\right)\nn\\
&+ C_{j,3}^{(i)}\left(s_{i,2} - C_{i,2}^{(j)}s_{j,1}\right).\nn
\end{align}
Then, $\mbH_{i,3}$ is given by
\begin{align}
\mbH_{i,3} = \begin{psmallmatrix} B_{j_1,3}H_{j_1} + C_{j_1,3}\mbH_{j_1,2} \\ \vdots \\ B_{j_{\pi_i},3}H_{j_{\pi_i}} + C_{j_{\pi_i},3}\mbH_{j_{\pi_i},2} \end{psmallmatrix} - \begin{psmallmatrix} C_{j_1,3}^{(i)}\mbH_{j_1,2}^{(i)} \\ \vdots \\ C_{j_{\pi_i},3}^{(i)}\mbH_{j_{\pi_i},2}^{(i)} \end{psmallmatrix}.\label{eq:h3}
\end{align}
Correspondingly, we have
\begin{align}
G_{i,3} = \begin{psmallmatrix} B_{j_1,3}\mSig_{n_{j_1}}B_{j_1,3}' + C_{j_1,3}\mbG_{j_1,2}C_{j_1,3}' \\ \vdots \\ B_{j_{\pi_i},3}\mSig_{n_{j_{\pi_i}}}B_{j_{\pi_i},3}' + C_{j_{\pi_i},3}\mbG_{j_{\pi_i},2}C_{j_{\pi_i},3}' \end{psmallmatrix}- \begin{psmallmatrix} C_{j_1,3}^{(i)}G_{j_1,2}^{(i)}(C_{j_1,3}^{(i)})' \\ \vdots \\ C_{j_{\pi_i},3}^{(i)}G_{j_{\pi_i},2}^{(i)}(C_{j_{\pi_i},3}^{(i)})' \end{psmallmatrix}.\label{eq:g3}
\end{align}
Therefore, the oracle action can be written as
\[
u_{i,4}^o = A_{i,4}u_{i,3}^o + B_{i,4}y_{i,4} + C_{i,4}(r_{i,4} - D_{i,3}s_{i,2} + \mT_{i,3}r_{i,2}),
\]
where $A_{i,4},B_{i,4},C_{i,4}, D_{i,3}$ are defined accordingly and
\[
\mT_{i,3} := \begin{psmallmatrix} C_{j_1,3}^{(i)} C_{i,2}^{(j_1)} &  & \\  & \ddots & \\ &  & C_{j_{\pi_i},3}^{(i)}C_{i,2}^{(j_{\pi_i})} \end{psmallmatrix}.
\]

Following identical steps, for $t \geq 1$, the OEDOL algorithm is given by
\begin{align}
&u_{i,t}^o = \mA_{i,t}u_{i,t-1}^o + \mB_{i,t} y_{i,t} + \mC_{i,t} w_{i,t},\label{eq:x_iter}\\
&\mhSig_{i,t} = \mA_{i,t}\mhSig_{i,t-1},\label{eq:sig_iter}
\end{align}
where $w_{i,t}$ is the innovation term extracted from the received information, which evolves according to
\begin{equation}
w_{i,t} = r_{i,t} - \mD_{i,t-1} s_{i,t-2} + \mT_{i,t-1} w_{i,t-2}.
\end{equation}
The weighting matrices $\mA_{i,t}$, $\mB_{i,t}$, $\mC_{i,t}$, $\mD_{i,t}$, and $\mT_{i,t}$ are defined by
\begin{align}
&\begin{bmatrix} \mB_{i,t} & \mC_{i,t} \end{bmatrix} =\mhSig_{i,t-1}\mtH_{i,t-1}^T\times\nn\\
&\hspace{.9in}\left(\mtH_{i,t-1}\mhSig_{i,t-1}\mtH_{i,t-1}^T + \mtG_{i,t-1}\right)^{-1},\label{eq:bc}\\
&\mA_{i,t} = \eye - \mB_{i,t}\mH_i - \mC_{i,t}\mbH_{i,t-1},\label{eq:a}\\
&\mD_{i,t} = \mathrm{col}\Big\{\mC_{j_1,t}^{(i)},\cdots,\mC_{j_{\pi_i},t}^{(i)}\Big\}\label{eq:d}\\
&\mT_{i,t} = \begin{psmallmatrix} \mC_{j_1,t}^{(i)}\mC_{i,t-1}^{(j_1)} & \cdots & \mZero \\ \vdots & \ddots & \vdots \\ \mZero & \cdots & \mC_{j_{\pi_i},t}^{(i)}\mC_{i,t-1}^{(j_{\pi_i})}\end{psmallmatrix}.\label{eq:t}
\end{align}
where $\mtH_{i,t} = \mathrm{col}\{\mH_i,\mbH_{i,t}\}$, $\mtG_{i,t} = \mathrm{diag}\{\mSig_{n_i},\mbG_{i,t}\}$ and $\mbG_{i,t} = \mathrm{diag}\{\mG_{i,t}\}$. By \eqref{eq:h3} and \eqref{eq:g3}, the intermediate parameters $\mbH_{i,t}$ and $\mG_{i,t}$ evolve according to
\begin{align}
&\mbH_{i,t} = \begin{psmallmatrix} \mB_{j_1,t}\mH_{j_1} + \mC_{j_1,t}\mbH_{j_1,t-1} \\ \vdots \\ \mB_{j_{\pi_i},t}\mH_{j_{\pi_i}} + \mC_{j_{\pi_i},t}\mbH_{j_{\pi_i},t-1} \end{psmallmatrix} - \begin{psmallmatrix}\mC_{j_1,t}^{(i)}\mbH_{j_1,t-1}^{(i)} \\ \vdots \\ \mC_{j_{\pi_i},t}^{(i)}\mbH_{j_{\pi_i},t-1}^{(i)}\end{psmallmatrix},\label{eq:h}\\
&\mG_{i,t} = \begin{psmallmatrix} \mB_{j_1,t}\mSig_{n_{j_1}}\mB_{j_1,t}^T + \mC_{j_1,t}\mbG_{j_1,t-1}\mC_{j_1,t}^T \\ \vdots \\ \mB_{j_{\pi_i},t}\mSig_{n_{j_{\pi_i}}}\mB_{j_{\pi_i},t}^T + \mC_{j_{\pi_i},t}\mbG_{j_{\pi_i},t-1}\mC_{j_{\pi_i},t}^T\end{psmallmatrix} - \begin{psmallmatrix} \mC_{j_1,t}^{(i)}\mG_{j_1,t-1}^{(i)}\left(\mC_{j_1,t}^{(i)}\right)^T \\ \vdots \\\mC_{j_{\pi_i},t}^{(i)}\mG_{j_{\pi_i},t-1}^{(i)}\left(\mC_{j_{\pi_i},t}^{(i)}\right)^T\end{psmallmatrix}\label{eq:g}
\end{align}
and we initialize the parameters as $\mbH_{j,\tau} = \vec{0}$ and $\mG_{j,\tau} = \vec{0}$ for $\tau < 1$.
Then, agent-$i$ sends
\[
s_{i,t} = u_{i,t}^o - \mA_{i,t}u_{i,t-1}^o.
\]
The detailed description of the algorithm is provided in Table \ref{tab:oedol}.

\begin{table}[t!]
\renewcommand{\arraystretch}{1.25}
  \caption{The description of the OEDOL algorithm.}
  \centering
  \begin{tabularx}{\textwidth}{ l}
          \hline
          \textbf{Algorithm:} The OEDOL Algorithm \\
          \hline
          \textbf{Initialization:} \\
          For $i = 1$ to $m$ do \\
          \hspace{1em} $u_{i,0}^o = \bar{x}$, $\mhSig_{i,0} = \mSig_x$, \\
          \hspace{1em} $\mbH_{i,\tau} = \vec{0}$, $\mG_{i,\tau} = \vec{0}$, and $w_{i,\tau} = \vec{0}$ for $\tau < 1$\\
          End for\\
          \textbf{Iterations:}\\
          Do for $t \geq 1$\\
          \hspace{1em} For $i = 1$ to $m$ do \\
          \hspace{2em} \textbf{Construction of Weights:}\\
          \hspace{2em} For $j = 1$ to $m$ do \\
          \hspace{3em} \textit{Calculate $\mbH_{j,t}$ and $\mG_{j,t}$ by \eqref{eq:h} and \eqref{eq:g}.} \\
          \hspace{3em} \textit{Determine combination matrices via \eqref{eq:bc}, \eqref{eq:a}, \eqref{eq:d}, and \eqref{eq:t}.}\\
          \hspace{2em} End for\\
          \hspace{2em} \textit{Construct $r_{i,t}$ through received $s_{j,t-1}$ for $j \in \cN_i$} \\
          \hspace{2em} \textbf{Extraction of Innovation:}\\
          \hspace{2em} $w_{i,t} = r_{i,t} - \mD_{i,t-1}s_{i,t-2} + \mT_{i,t-1}w_{i,t-2}$\\
          \hspace{2em} \textbf{Update:}\\
          \hspace{2em} $u_{i,t}^o = \mA_{i,t}u_{i,t-1}^o + \mB_{i,t}y_{i,t} + \mC_{i,t}w_{i,t}$\\
          \hspace{2em} $\mhSig_{i,t} = \mA_{i,t}\mhSig_{i,t-1}$\\
          \hspace{2em} \textit{Disclose $s_{i,t} = u_{i,t}^o - \mA_{i,t}u_{i,t-1}^o$ to the neighbors.}\\
          \hspace{1em} End for\\
          \hline
        \end{tabularx}\label{tab:oedol}
\end{table}

\subsection{Computational Complexity}

In \eqref{eq:x_iter}, the combination matrices $\mA_{i,t}, \mB_{i,t}$, $\mC_{i,t}$, $\mD_{i,t-1}$, and $\mT_{i,t-1}$ are independent of the streaming data although they are time-varying. Hence they can be computed before-hand. In that case, the computational complexity of the iterations for each agent is dominated by the term $\mC_{i,t}w_{i,t}$. Therefore, the average computational complexity is on the order of $p^2\pi^2$, where $\pi^2 := 1/m \sum_{i=1}^m \pi_i^2$, i.e., $O(p^2\pi^2)$. Otherwise, the computational complexity of the algorithm is mainly dominated by the matrix inversion in \eqref{eq:bc}, note that $\mbG_{i,t} \in \mathbb{R}^{p\pi_i \times p\pi_i}$, unless the network is sparsely connected, i.e., $\pi_i \ll m$ for $i=1,\cdots,m$. Therefore, over a non-sparse network, the average complexity is on the order of $mp^3\pi^3$ ($\pi^3$ is defined accordingly) since each agent $i$ computes $B_{j,t}$ and $C_{j,t}$ for $j=1\cdots,m$. In particular, the complexity is given by $O\left(mp^3\pi^3\right)$, while it is $O\left((qm)^3\right)$ for the ODOL algorithm. Note that over tree networks, we have $m-1$ edges and correspondingly average neighborhood size is small. Hence, disclosure of local estimates over tree networks also reduces the computational complexity compared to the time-stamped disclosure strategy in general (in addition to the substantial reduction in communication). In Table \ref{tab2}, we tabulate the computational complexities of the introduced algorithms.

We point out that diffusion or consensus based algorithms have relatively low complexity, i.e., on the order of $p^2$ in the least-mean-square based algorithms and on the order of $p^3$ in quasi-Newton based algorithms, since exchanged information is handled irrespective of the content. Such algorithms also present appealing performance for certain applications in addition to the low computational complexity. However, they do not achieve the oracle performance.

\begin{table}[t!]
\renewcommand{\arraystretch}{1.25}
\caption{A comparison of the computational complexities of the proposed algorithms.}
\label{tab2}
\begin{center}
  \begin{tabular}{| c | c | c |}
    \hline
    Algorithm                  & Without weights & Pre-computed weights \\
    \hline
    ODOL algorithm           & $O\left((qm)^3\right)$   & $O\left((qm)^2\right)$ \\
    \hline
    OEDOL algorithm          & $O\left(mp^3\pi^3\right)$ &$O\left(p^2\pi^2\right)$ \\
    \hline
  \end{tabular}
\end{center}
\end{table}

In the following, we analyze whether the agents can still achieve the oracle performance through the exchange of local estimates over the networks not in tree structure.

\subsection{Tree Networks Involving Cell Structures}
While constructing the spanning tree, we cancel certain communication links in order to avoid multi-path information propagation. However, we also observe that in a fully connected network agents can achieve the oracle performance through the disclosure of local observations. In particular, since all of the agents are connected, each agent can receive the observations across the network directly. Correspondingly, in a fully connected network, we can achieve identical performance with the ODOL algorithm only through the disclosure of the local estimates as stated in the following corollary formally.

{\bf Corollary 3.1:}
{\em Consider the team-problem over a fully connected network, in which $r=p$. Then, exchange of local estimates can lead to the team-optimal solution, i.e., agents can achieve the oracle performance.}

{\em Proof:}
Over a fully connected network, $\kappa_i = 1$ and the oracle action is given by
\begin{equation}\label{eq:full}
u_{i,t}^o = E[\bx | \{\by_{i,\tau} = y_{i,\tau}\}^{\tau \leq t}, \{\by_{j,\tau} = \vy_{j,\tau}\}_{j \in \cN_i}^{\tau \leq t-1}\Bigg]
\end{equation}
and we can also obtain \eqref{eq:full} by \eqref{eq:x11} since we have
\[
\Delta_{j,i,t}^o = \delta_{j,t}^o \setminus \{\delta_{i,t}^o \setminus \{y_{i,t}\}\},
\]
which implies that $\Delta_{j,i,t}^o$ is constructible from $\delta_{i,\tau}^o$ and $\delta_{j,\tau}^o$ for $\tau \leq t$. The proof is concluded. \hfill $\square$

We point out that the team-optimal strategies can fail if a link or node failure occurs. However, once a link failure is detected, team-optimal strategies can be recomputed by eliminating the failed link in the new network configuration. Hence, through such strategies, we can increase the robustness of the team against link and node failures.

\begin{figure}
  \centering
  \includegraphics[width=2.5in]{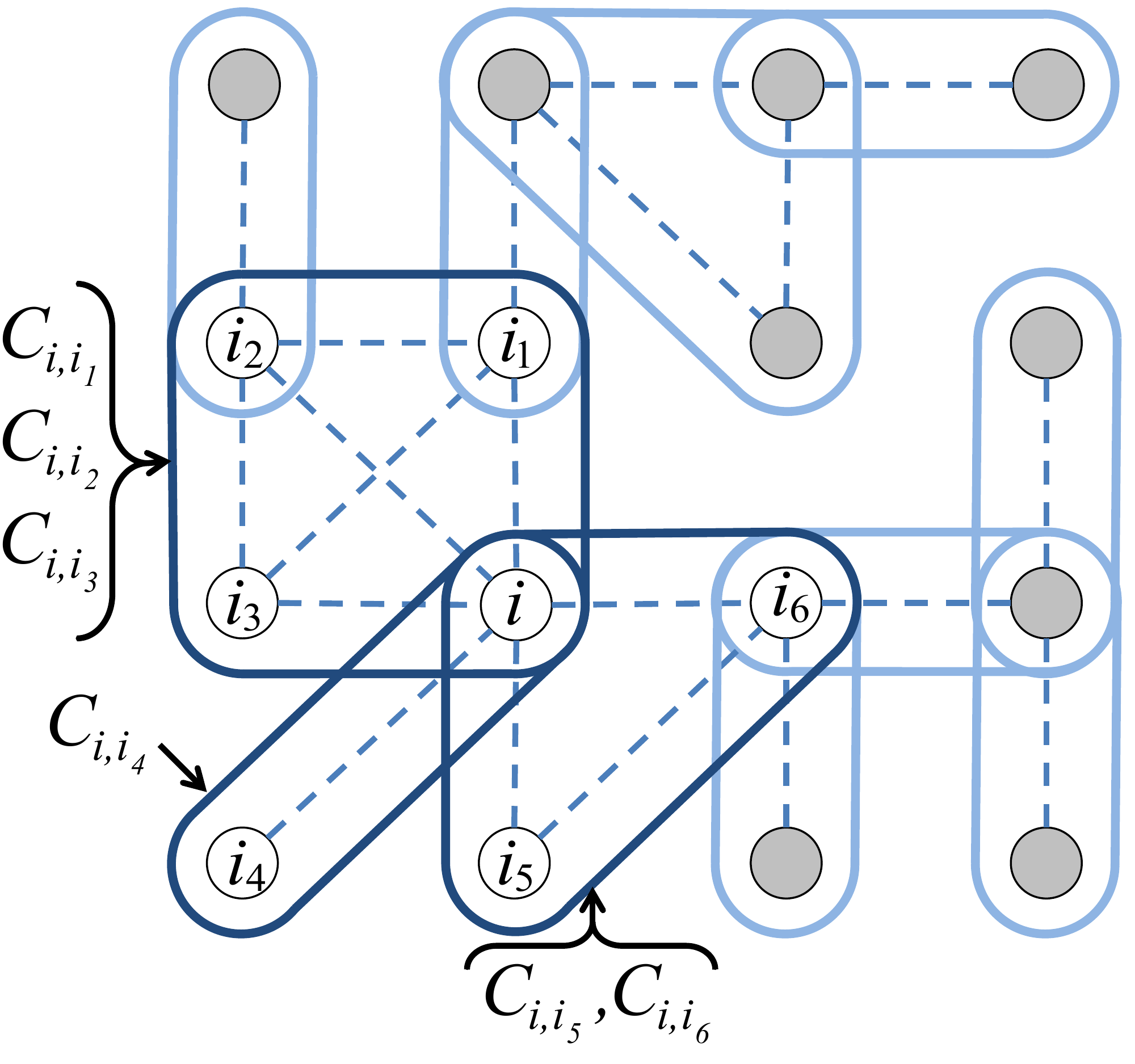}\\
  \caption{An example tree network involving cell structures.}\label{fig:cell}
\end{figure}

We define a ``cell structure" as a sub-network in which all agents are connected to each other. Intuitively, considering a cell structure as a ``single" agent, the cell (i.e., all the agents in the cell) can be involved in the tree such that the agents can still achieve the oracle performance through the disclosure of local estimates (although there may be loops in the cell). We list the features of the cell structures, e.g., seen in Fig. \ref{fig:cell}, as follows:

\begin{itemize}
\item Agents out of a cell can connect to at most one of the agents within that cell.
\item A cell structure consists of at least 2 agents.
\item An agent can belong to more than one cell.
\item Two different agents cannot belong to more than one cell at the same time.
\item All of the agents belong to at least a cell in a connected network.
\item Each agent has also the knowledge of the cells of the other agents.
\item Each agent labels its cells from its own and its first order neighbor's point of view. As an example, for agent-$i$, $\cC_{i,i_1}$ denotes the cell involving both $i$ and $i_1$. Note that if the same cell also includes $i_2$, $\cC_{i,i_1} = \cC_{i,i_2}$.
\end{itemize}

The following theorem shows that agents can achieve the oracle performance over tree networks involving cell structures through the disclosure of the local estimates.

{\bf Theorem 3.2:}
{\em Consider the team problem over tree networks involving cell structures. Then, exchange of local estimates can lead to the team-optimal solution, i.e., agents can achieve the oracle performance.}

{\em Proof:}
Initially, we have $\Delta_{j,i,1}^o = \delta_{j,1} = \{y_{j,1}\}$ and the oracle action is also given by \eqref{eq:x11} over this network topology. Note that the information received by $j$ at $t = 2$ is given by
\(
\delta_{j,2}^o = \Bigg\{y_{j,2},\Big\{\overbrace{y_{k,1}}^{=\Delta_{k,j,0}^o}\Big\}_{k \in \cN_j},\delta_{j,1}^o\Bigg\},
\)
which yields
\begin{align}
\Delta_{j,i,2}^o &= \delta_{j,2}^o \setminus \left\{\delta_{j,1}^o \cup \bigcup_{k \in \cC_{i,j}}\left\{y_{k,1}\right\}\right\},\nn\\
&=\delta_{j,2}^o \setminus \left\{\delta_{j,1}^o \cup \bigcup_{k \in \cC_{i,j}}\Delta_{k,j,1}^o\right\}\nn
\end{align}
and $\Delta_{j,j,t}^o = \varnothing$ by definition. Due to the cell structure, we have
\begin{align}
\bigcup_{k \in \cC_{j,i}}\Delta_{k,j,1}^o &= \Delta_{i,j,1}^o\cup\bigcup_{k \in \cC_{i,j} \setminus j} \Delta_{k,i,1}^o,\nn\\
&= \delta_{i,1}^o \cup \bigcup_{k \in \cC_{i,j}\setminus j} \delta_{k,1}^o.\nn
\end{align}
Correspondingly, for $t > 0$ we have
\begin{equation}
\Delta_{j,i,t}^o = \delta_{j,t}^o \setminus \left\{\delta_{j,t-1}^o \cup \Delta_{i,j,t-1}^o\cup \bigcup_{k \in \cC_{i,j}\setminus j}\Delta_{k,i,t-1}^o\right\}\label{eq:cell}
\end{equation}
and $\Delta_{j,i,t}^o$ is constructible by the sets $\delta_{i,\tau}^o$ and $\delta_{j,\tau}^o$ for $j \in \cN_i$ and $\tau \leq t$. Note that over tree networks, $\cC_{i,j} \setminus j = \varnothing$ for $i=1,\cdots,m$, $j\in\cN_i$, and \eqref{eq:cell} leads to \eqref{eq:4}. Hence, for $t > 0$ we obtain \eqref{eq:full3} and the proof is concluded. \hfill $\square$

Note that the network can have loops within the cell structures and agents can still achieve the oracle performance through the diffusion of the local estimates. This increases the robustness of the team strategies against the link failures. In the sequel, we provide the sub-optimal extensions of the algorithms for practical applications.

\begin{figure}[t]
  \centering
  \includegraphics[width=3.5in]{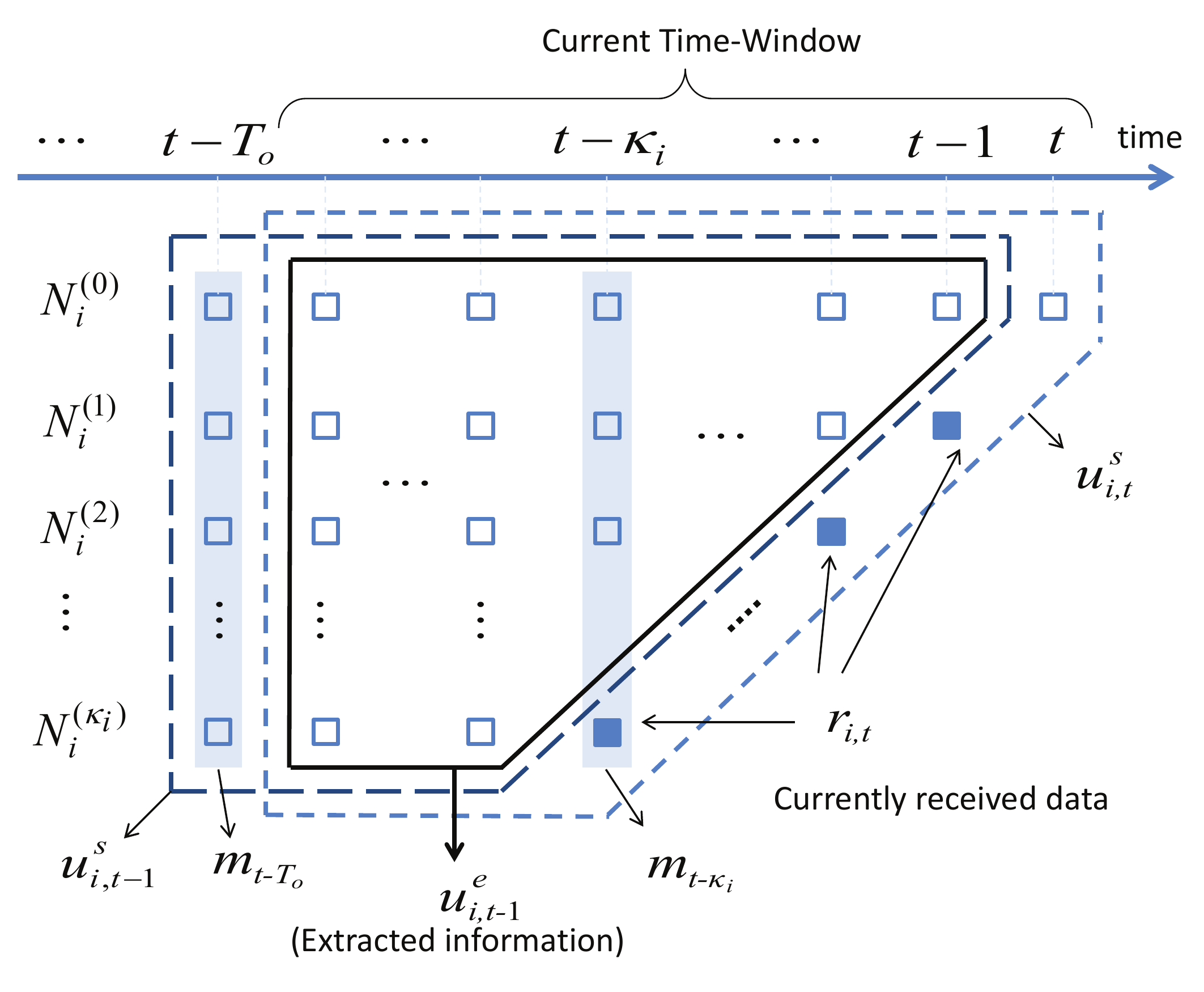}\\
  \caption{Information aggregation illustration. Small squares represent the information across the agents in the corresponding neighborhood and time.}\label{fig:ia}
\end{figure}

\section{Sub-optimal Approaches}
Minimization of the cost function \eqref{eq:costlow2} optimally requires relatively excessive computations. We aim to mitigate the problem sub-optimally yet in a computationally efficient approach while achieving comparable performance with the optimal case. As an example, we can approximate the cost measure \eqref{eq:costlow2} through $T_o \geq \max\{\kappa_i\}_{i=1,\cdots,m}$ size time-windowing as follows
\begin{equation}
\sum_{t=1}^T\min\limits_{\substack{\eta_{i,t}\in\Omega^s_{i,t}}} E\|\bx - \eta_{i,t}(\delta_{i,t}^s)\|^2,\label{eq:costlows}
\end{equation}
where agent-$i$ has the information set
\begin{equation}\label{eq:deltas}
\delta_{i,t}^s :=
\left\{\left\{y_{i,\tau}\right\}^{t-T_o < \tau \leq t},\left\{y_{j,\tau}\right\}^{t-T_o < \tau \leq t-1}_{j\in \cN_i^{(1)}},\cdots,\left\{y_{j,\tau}\right\}^{t-T_o < \tau \leq t-\kappa_i}_{j\in \cN_{i}^{(\kappa_i)}}\right\},
\end{equation}
if $t \geq T_o$ and other cases are defined accordingly, $\sigma_{i,t}^s$ is the sigma-algebra generated by $\delta_{i,t}^s$, and $\Omega_{i,t}^o$ denotes the set of all $\sigma_{i,t}^s$-measurable functions from $\mathbb{R}^{q\kappa}\times\mathbb{R}^{q\pi_i(\kappa-1)}\times\cdots \times \mathbb{R}^{q\pi_i^{\kappa_i}(\kappa-\kappa_i)}$ to $\mathbb{R}^p$. Then, team-optimal action is given by
\[
u_{i,t}^s = E[\bx | \bdel_{i,t}^s = \delta_{i,t}^s],
\]
where $\bdel_{i,t}^s$ is the set of corresponding random parameters.

In Fig. \ref{fig:ia}, we illustrate the time-windowing approach for time stamped information disclosure. We define a memory element $m_t$ denoting the expectation of the state conditioned on all measurement at time $t$, i.e.,
\(
m_t \defi E[\bx | \{\by_{i,t} = y_{i,t}\}_{i=1}^m].
\)
We consider that $\bar{x} = \mZero$. Then, we can write
\(
m_t = \mM y_{t},
\)
where $\mM \defi \mSig_x \mH' \left(\mH\mSig_x\mH' + \mSig_n\right)^{-1}$ and $y_t \defi \mathrm{col}\{y_{1,t}, \cdots,y_{m,t}\}$.

Let $u_{i,t-1}^{e}$ represent the extracted information from the estimate $u_{i,t-1}^s$ via the memory element $m_{t-T_o}$, i.e.,
\begin{align}
u_{i,t-1}^{e} \defi E\left[\bx \big|\{\by_{i,\tau} = y_{i,\tau}\}^{t-T_o+1 < \tau \leq t-1},\cdots,\{\by_{j,\tau} = y_{j,\tau}\}^{t - T_o + 1 < \tau \leq t-\kappa_i-1}_{j\in \cN_{i}^{(\kappa_i)}}\right]\nn
\end{align}
if $t \geq T_o$ and other cases are defined accordingly. Then, we can write $u_{i,t-1}^{e} = \mN_i y_{i,t-1}^{e}$, where $\mN_i = \mSig_x\mhH_i'\left(\mhH_i\mSig_x\mhH_i' + \mhSig_{n_i}\right)^{-1}$ and $y_{i,t-1}^{e}$ is the vector collecting the measurements within that window (See Fig. \ref{fig:ia}).
Additionally, we define
\[
\mhH_{i} \defi \mathrm{col} \left\{\mH_i, \mH_i,\mH_i^{(1)},\mH_i,\mH_i^{(1)},\mH_i^{(2)},\cdots,\mH_i,\mH_i^{(1)},\cdots,\mH_i^{(\kappa_i)}\right\}
\]
and $\mH_i^{(k)} \defi \mathrm{col}\left\{\mH_j\right\}_{j\in\cN_{i}^{(k)}}$. Correspondingly, we have
\[
\mhSig_{n_i} \defi \mathrm{diag} \left\{\mSig_{n_i}, \mSig_{n_i},\mSig_{n_i}^{(1)},\mSig_{n_i},\mSig_{n_i}^{(1)},\mSig_{n,i}^{(2)},\cdots,\mSig_{n_i},\mSig_{n_i}^{(1)},\cdots,\mSig_{n_i}^{(\kappa_i)}\right\}
\]
and $\mSig_{n_i}^{(k)} \defi \mathrm{diag}\left\{\mSig_{n_{j}}\right\}_{j\in\cN_{i}^{(k)}}$. Therefore, for $t \geq T_o$, we have
\[
u_{i,t-1}^s = E[\bx | \bu_{i,t-1}^e = u_{i,t-1}^e,\bm_{t-T_o} = m_{t-T_o}].
\]
Note that $E[\bx|\bm_{t-T_o} = m_{t-T_o}] = m_{t-T_o}$ and $\mhSig := E[(\bx - m_{t-T_o})(\bx - m_{t-T_o})'] = \mSig_x H'(H\mSig H' + \mSig_n)^{-1}H\mSig = (\eye - MH)\mSig$. Then, the distribution of $\bx$ and $\bu_{i,t-1}^e$ conditioned on $\bm_{t-T_o} = m_{t-T_o}$ is given by
\[
\begin{psmallmatrix} \bx \\ \bu_{i,t-1}^e\end{psmallmatrix} | \bm_{t-T_o} = m_{t-T_o} \sim \mathbb{N} \left(\begin{psmallmatrix} m_{t-T_o} \\ N_i\mhH_i m_{t-T_o} \end{psmallmatrix},\begin{psmallmatrix} \mhSig & \mhSig \mhH_i'N_i' \\ N_i \mhH_i \mhSig\; & \;N_i \mhH_i \mhSig \mhH_i' N_i' + N_i \mhSig_{n_i}N_i' \end{psmallmatrix}\right)
\]
and we obtain
\begin{align}
u_{i,t-1}^s &= m_{t-T_o} + \mhSig \mhH_i' N_i' (N_i \mhH_i \mhSig \mhH_i' N_i' + N_i \mhSig_{n_i}N_i')^{-1}(u_{i,t-1}^e - N_i \mH_i m_{t-T_o})\nn\\
&= K_i m_{t-T_o} + L_i u_{i,t-1}^e,\label{eq:us}
\end{align}
where
\begin{align}
&L_i := \mhSig \mhH_i' N_i' (N_i \mhH_i \mhSig \mhH_i' N_i' + N_i \mhSig_{n_i}N_i')^{-1},\nn\\
&K_i := \eye - L_iN_i\mhH_i.\nn
\end{align}

Next, we aim to compute $u_{i,t}^s$ through $u_{i,t-1}^s$ and $\delta_{i,t}^s = \{y_{i,t-1}^e, y_{i,t},r_{i,t}\}$, where $r_{i,t}$ is a vector consisting of currently received measurements as seen in Fig. \ref{fig:ia}. Let $\mbH_{i} := \col\{H_i,H_i^{(1)},\cdots,H_i^{(\kappa_i)}\}$ and $\mbSig_i := \col\{\mSig_{n_i},\mSig_{n_i}^{(1)},\cdots,\mSig_{n_i}^{(\kappa_i)}\}$. Since $E[(\bx - u_{i,t-1}^e)(\bx - u_{i,t-1}^e)'] = \mSig_x \mhH_i'(\mhH_i\mSig_x \mhH_i' + \mhSig_n)^{-1}\mhH_i\mSig_x = (\eye - N_i\mhH_i)\mSig_x$, we have
\[
u_{i,t}^s = u_{i,t-1}^e + \mSig_{x,i}\mbH_i'(\mbH_i\mSig_{x,i}\mbH_i' + \mbSig_{n_i})^{-1}\left(\begin{psmallmatrix}y_{i,t}\\ r_{i,t}\end{psmallmatrix} - \mbH_i u_{i,t-1}^e\right),
\]
where $\mSig_{x,i} := (\eye - N_i\mbH_i)\mSig_x$. By \eqref{eq:us}, $u_{i,t-1}^e = L_i^{-1}(u_{i,t-1}^s - K_i m_{t-T_o})$, and we define the combination matrices
\begin{align}
&\begin{bmatrix} \mB_i & \mC_i \end{bmatrix} = \mSig_{x,i}\mbH_i^T \left(\mbH_i\mSig_{x,i}\mbH_i^T + \mbSig_{n_i} \right)^{-1},\label{eq:bb}\\
&\mA_i \defi \left(\eye - \begin{bmatrix} \mB_i & \mC_i \end{bmatrix} \mbH_i\right)\mL_i^{-1},\label{eq:aa}\\
&\mD_i \defi \mA_i\mK_i\mM. \label{eq:dd}
\end{align}
Therefore, for $t\geq T_o$, the sub-optimal distributed online learning (SDOL) algorithm is given by
\begin{equation}\label{eq:update}
u_{i,t}^s = \mA_i u_{i,t-1}^s + \mB_i y_{i,t} + \mC_i r_{i,t} - \mD_i y_{t-T_o}.
\end{equation}
In the following we provide several remarks about the practical implementation of the SDOL algorithm.\\

{\bf Remark 4.1:}
{\em \begin{itemize}
\item We point out that in the update \eqref{eq:update} the matrices $\mA_i$, $\mB_i$, $\mC_i$, and $\mD_i$ are independent from the data and they are time invariant for $t\geq T_o$. Hence, they can be pre-computed and installed onto the agents. Then, the SDOL algorithm basically takes the linear average of the previous estimate, the current measurement and received data, and measurements at time $t-T_o$.

\item Different from the conventional approaches, the SDOL algorithm requires to memorize previous observations. Note that if $q < p/m$, storing measurements individually rather than a linear combination, i.e., $m_t$, might be more efficient in terms of memory usage.

\item The computational complexity of the SDOL algorithm is $O(p^2 + 2pmq)$ (or say $O(p^2)$ for $q \ll p/m$) due to matrix-vector multiplications in \eqref{eq:update}, and the algorithm requires disclosure of a data vector with $(m-1)\times q$ dimensions.

\item Finally, even though agent-$i$ uses update \eqref{eq:update} for $t<T_o$ assuming that\footnote{The action is no more team-optimal because of the assumption.} $y_{j,\tau} = 0$ for $j=1,\cdots,m$ and $\tau < 1$, we have
\begin{align}
E[(x-u_{i,\tau}^s)'(x-u_{i,\tau}^s)] &=  \mSig_{x,i} - \mSig_{x,i}\mbH_i^T \left(\mbH_i\mSig_{x,i}\mbH_i^T + \mbSig_{n_i} \right)^{-1}\mbH_i \mSig_{x,i}\nn\\
&= A_i L_i \mSig_{x,i}
\end{align}
for $\tau \geq T_o$ since at each iteration, agent-$i$ excludes the impact of $y_{t-T_o}$ on $u_{i,t}^s$. This sharp change results in as a breaking point on the learning curve as seen in Fig. \ref{fig:sdol}.
\end{itemize}}

In Table \ref{tab:sub}, we tabulate a detailed description of the SDOL algorithm. Note that SDOL algorithm is based on the aggregation of information. Correspondingly, we can apply the time-windowing approach to the disclosure of local estimates and formulate a sub-optimal efficient distributed online learning algorithm (SEDOL) over tree networks.

\begin{table}[t!]
\renewcommand{\arraystretch}{1.25}
  \caption{The description of the SDOL-$T_o$ algorithm.}
  \centering
  \begin{tabularx}{\textwidth}{ l}
          \hline
          \textbf{Algorithm:} The SDOL-$T_o$ algorithm \\
          \hline
          \textbf{Initialization:} \\
          \hspace{1em} $u_{i,0}^s = \vec{0}$ for all $i$\\
          \hspace{1em} $y_{i,\tau} = \vec{0}$ for $\tau < 1$ and all $i$\\
          \hspace{1em} \textit{Calculate combination matrices via \eqref{eq:bb}, \eqref{eq:aa} and \eqref{eq:dd}}.\\
          \textbf{Update:}\\
          Do for $t \geq 1$\\
          \hspace{1em} For $i = 1$ to $m$ do \\
          \hspace{2em} \textit{Construct $r_{i,t}$ through received $s_{j,t-1}$ for $j \in \cN_i$.}\\
          \hspace{2em} \textit{Store $r_{i,t}$ and $y_{i,t}$ in the memory accordingly.}\\
          \hspace{2em} $u_{i,t}^s = \mA_iu_{i,t-1}^s + \mB_{i}y_{i,t} + \mC_{i}r_{i,t} - \mD_iy_{t-T_o}$.\\
          \hspace{2em} \textit{Erase $y_{t-T_o}$ from the memory.} \\
          \hspace{2em} \textit{Diffuse $s_{i,t}$ to the neighboring agents.} \\
          \hspace{1em} End for\\
          \hline
        \end{tabularx}\label{tab:sub}
\end{table}

In the sequel, we also provide illustrative simulations showing the enhanced tracking performance due to the proposed algorithm over several distributed network scenarios.

\begin{figure}[t!]
  \centering
  \subfloat[Fully connected network]{\includegraphics[width=0.5\textwidth]{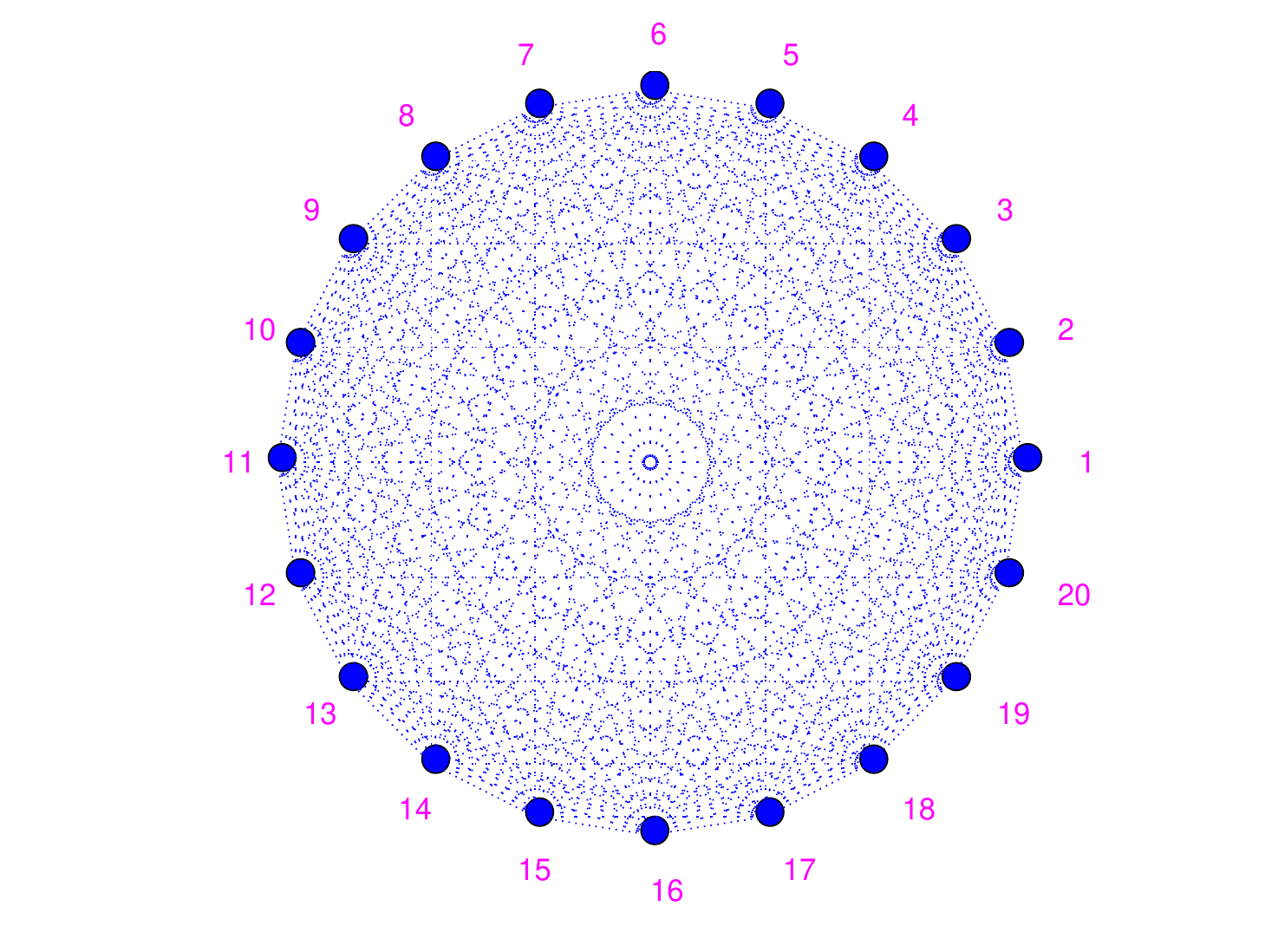}\label{fig:n_f}}
  \hfil
  \subfloat[Star network]{\includegraphics[width=0.5\textwidth]{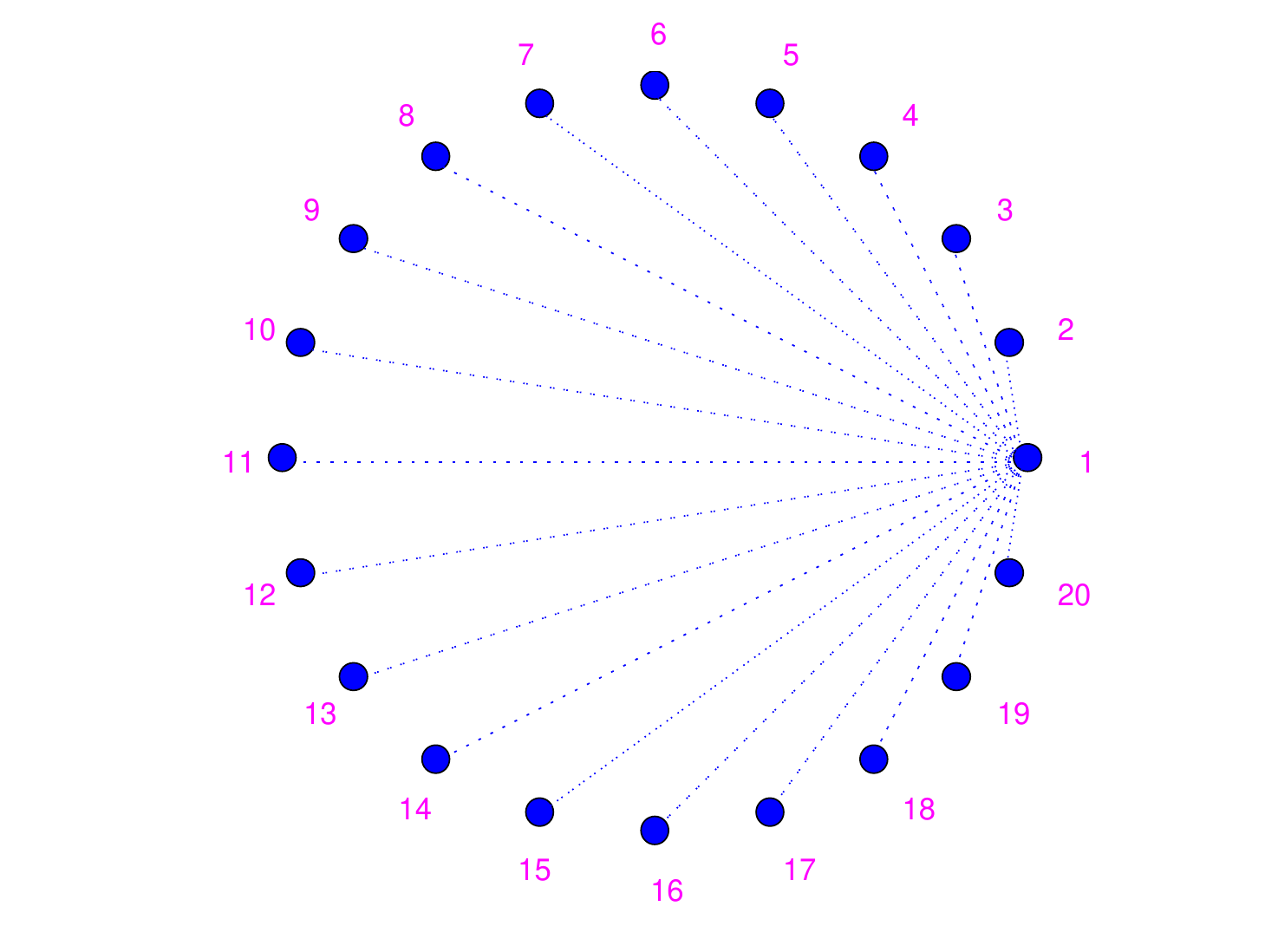}\label{fig:n_s}}\\
  \subfloat[Arbitrary network]{\includegraphics[width=0.5\textwidth]{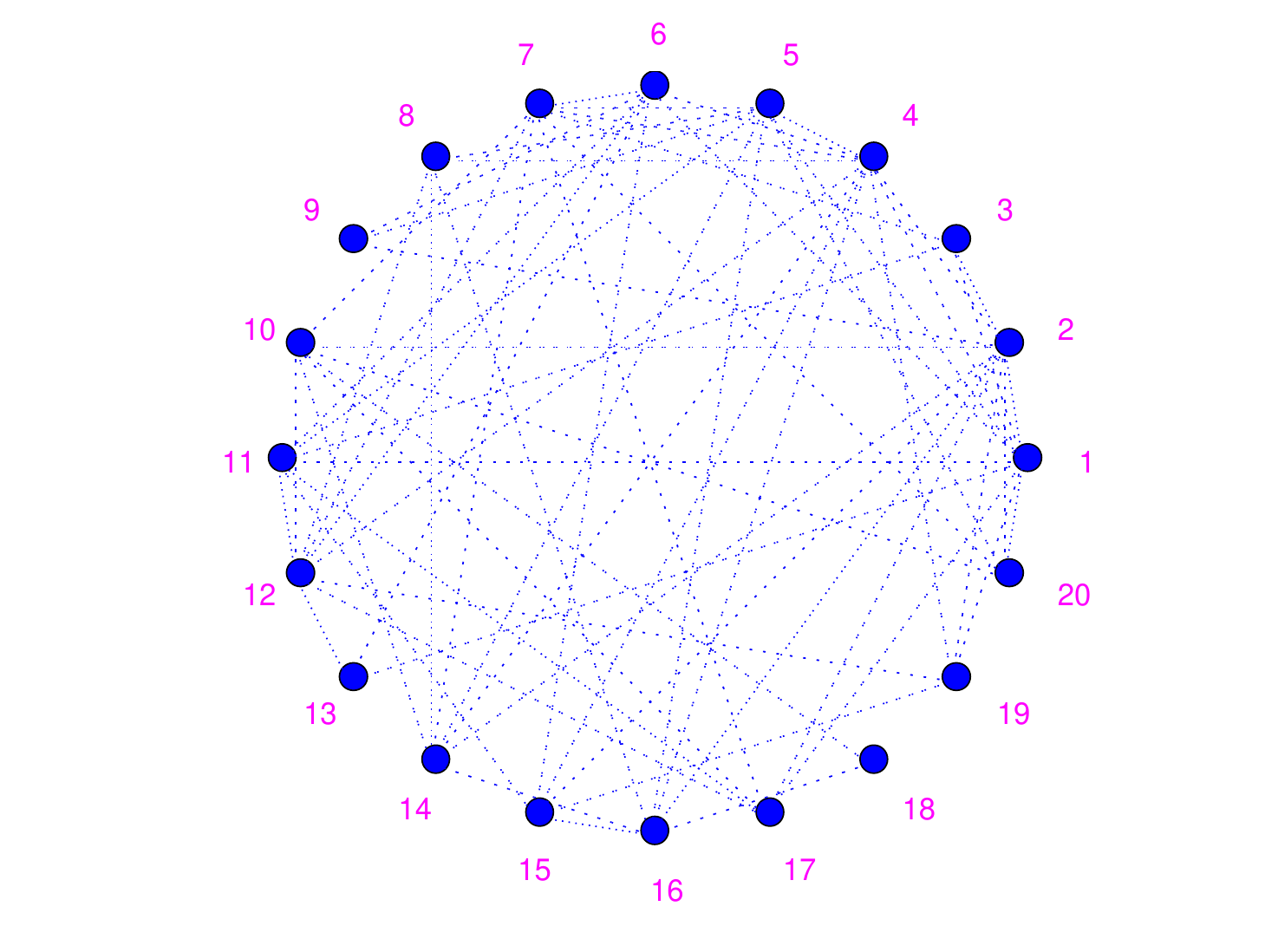}\label{fig:n_a}}
  \hfil
  \subfloat[Line network]{\includegraphics[width=0.5\textwidth]{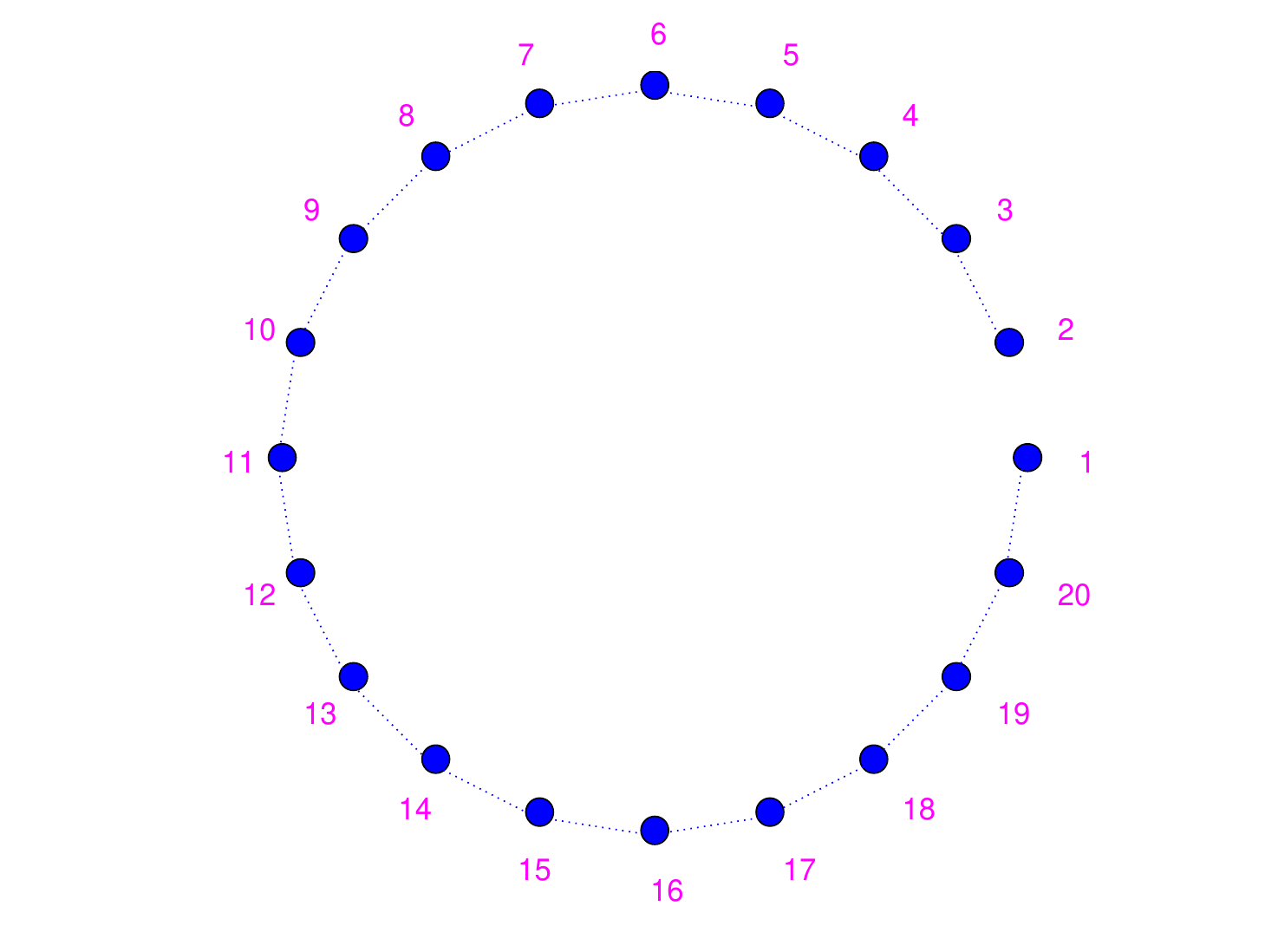}\label{fig:n_l}}
  \caption{Different network configurations.}
  \label{fig:networks}
\end{figure}

\section{Illustrative Examples}
In this section, we examine the performance of team optimal actions under different scenarios through numerical examples. To this end, we consider three main network structures, namely fully connected, star, and line networks with $m = 20$ agents, and a randomly generated network seen in Fig. \ref{fig:networks}. In a fully connected network, each agent has a connection with each other and any new measurement can be collected at each other after just one hop. This is the best possible connection such that the corresponding achievable team cost is the smallest. In a star network, there exists a pseudo-centralized agent, e.g., agent-$1$ in Fig. \ref{fig:n_s}, and all the other agents are connected to that agent. Therefore, the furthest distance between any two agents is $2$, e.g., there exist $2$-hops between agent-$2$ and agent-$3$ (agent-$2$ to agent-$1$ then agent-$1$ to agent-$3$) in Fig. \ref{fig:n_s}. We have generated the network in Fig. \ref{fig:n_a} randomly such that number of neighbors of agent-$i$, $\pi_i$, is chosen uniformly between $2$ to $10$. Therefore, the generated network can have loops, i.e., may not be a tree. Furthermore, in a line network, the connections between the agents form a line as seen in Fig. \ref{fig:n_l}. Contrary to the fully connected network, a line network has the least possible connection such that any two agents are connected via a certain number of hops. Note that there exist $m-1 = 19$ hops between agent-$1$ and agent-$2$ in Fig. \ref{fig:n_l}.

\begin{figure}
  \centering
  \includegraphics[width=0.7\textwidth]{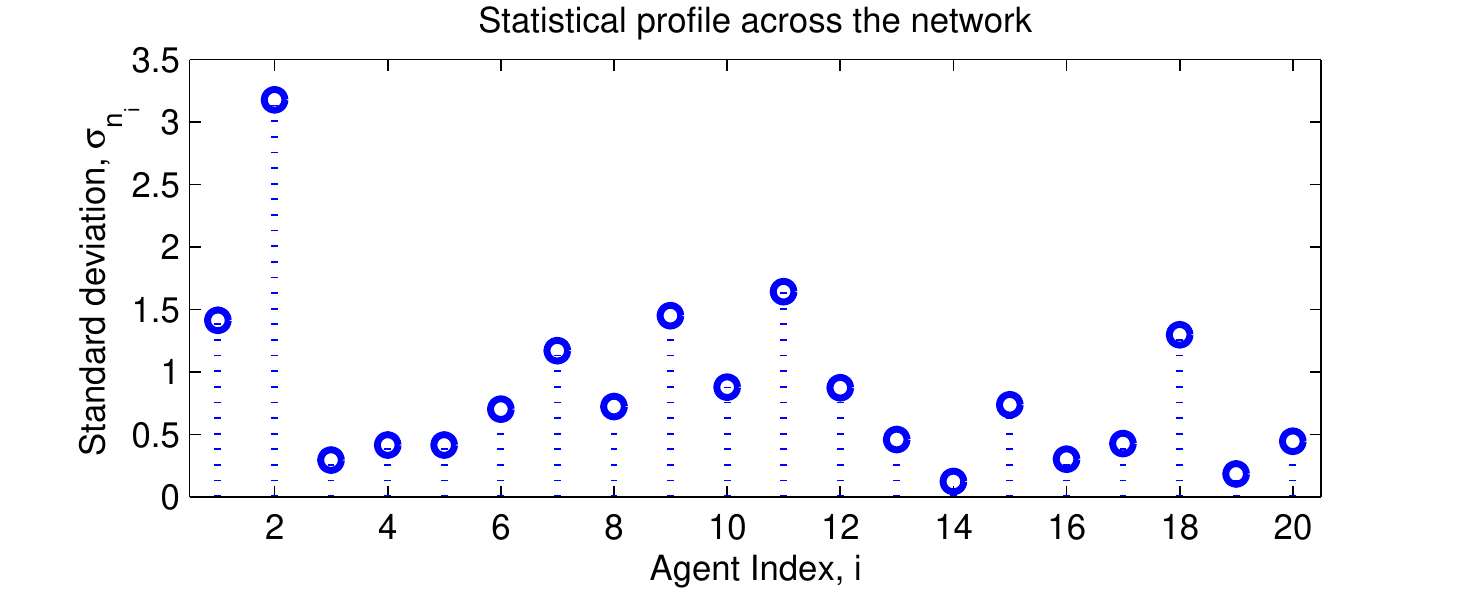}\\
  \caption{Standard deviation of measurement noise across the network.}\label{fig:noises}
\end{figure}

We analyze the performance of a team of agents over those network structures. We set $p=10$, $q=1$, $\bar{x} = \vec{0}$, $\mSig_x = \eye$, and the standard deviation of the zero-mean measurement noise is chosen randomly from a folded normal distribution as seen in Fig. \ref{fig:noises}. Furthermore, we choose $H_i$ randomly from normal distribution, $\mathbb{N}(\vec{0},\eye)$. In order to observe the change of oracle cost ($r \rightarrow \infty$) for different time horizons, we introduce
\begin{equation}
J(T) = \min\limits_{\substack{\eta_{i,t} \in \Omega_{i,t},\\ i = 1,\cdots,m, t = 1,\cdots,T}} \sum_{t=1}^T\sum_{i=1}^m E\|\bx - \eta_{i,t}(\delta_{i,t}^o)\|^2,\label{eq:Jt}
\end{equation}
which is a function of the length of time horizon, $T$. We re-emphasize that in \eqref{eq:Jt}, each action at $t=1,\cdots,T$ has a cumulative impact on the cost. Let $\eta_{i,t}^o$, for $i=1,\cdots,m$ and $t=1,\cdots,T$, be the oracle strategies for $J(T)$. Then, we introduce a terminal cost function defined by
\begin{equation}\label{eq:Pt}
P(T) = \sum_{i=1}^m E\|\bx - \eta_{i,T}^o(\delta_{i,T}^o)\|^2,
\end{equation}
which is the impact of final action on $J(T)$.

\begin{figure}[t!]
  \centering
  \includegraphics[width=0.9\textwidth]{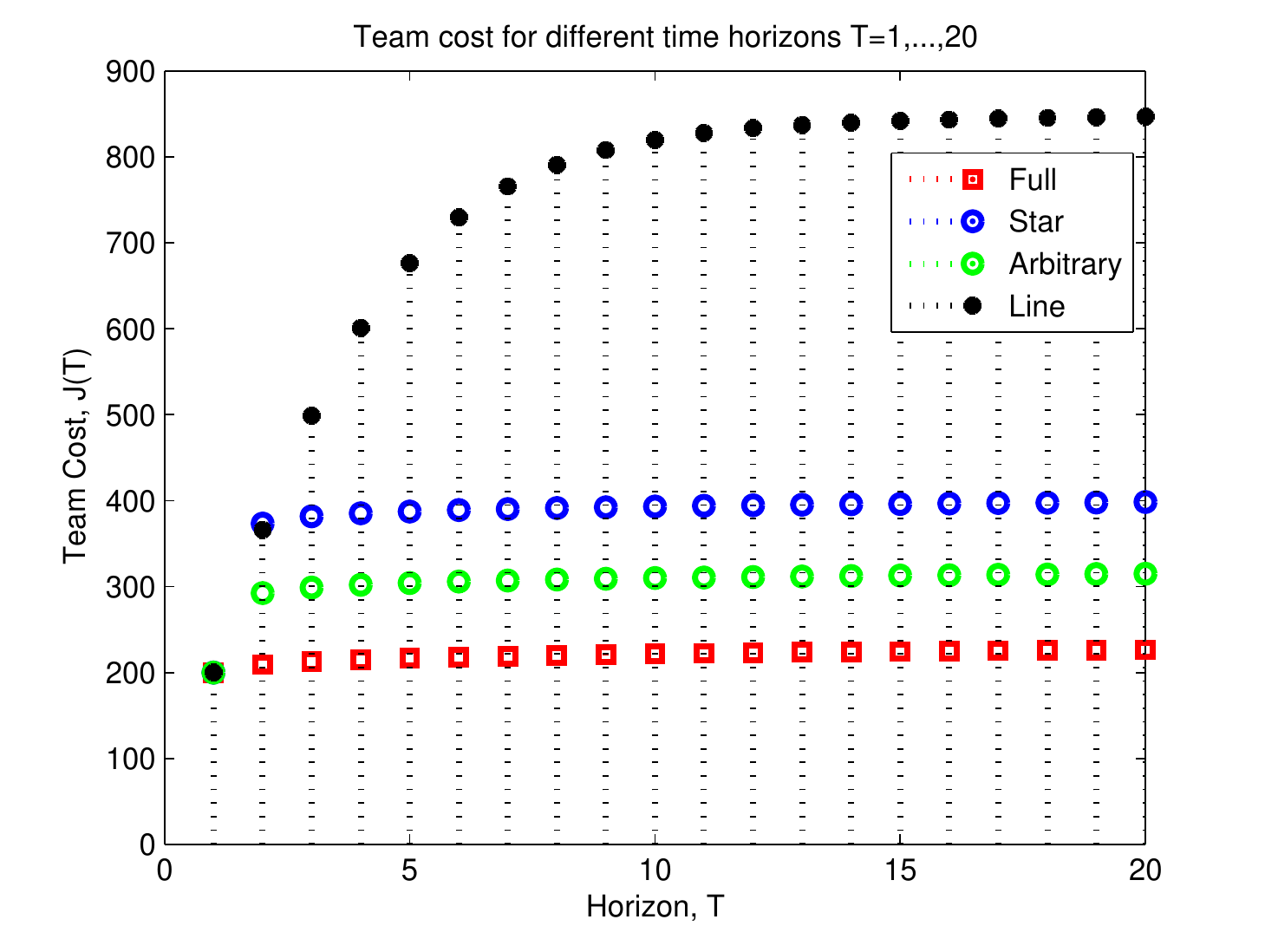}\\
  \caption{Comparison of oracle cost for different time horizons over certain networks.}\label{fig:comp_team}
\end{figure}

\begin{figure}[t!]
  \centering
  \includegraphics[width=0.9\textwidth]{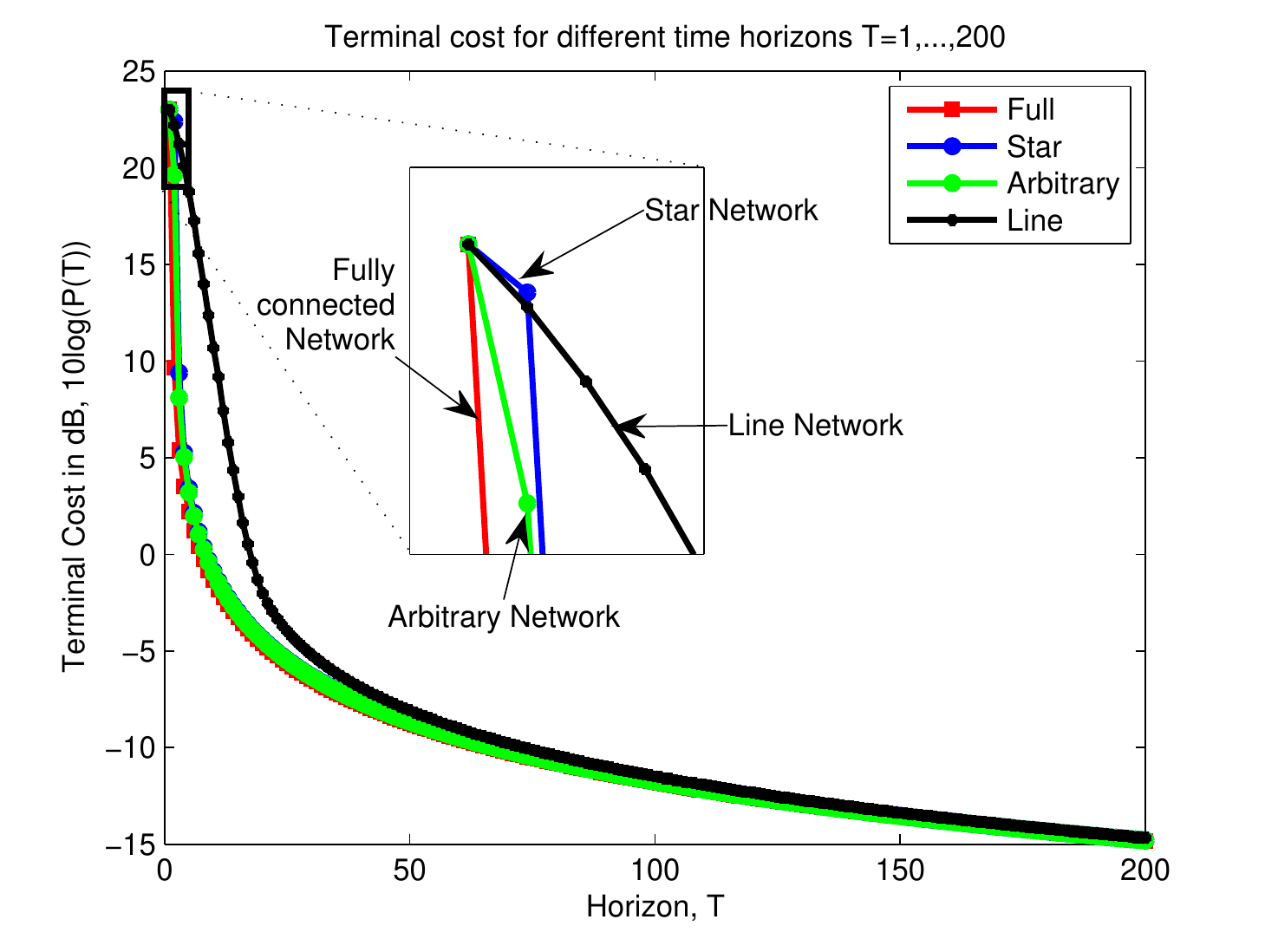}\\
  \caption{Comparison of terminal cost for different time horizons over certain networks.}\label{fig:comp_terminal}
\end{figure}

In Fig. \ref{fig:comp_team}, we compare $J(T)$, for $T=1,\cdots,20$, over the fully connected, star, arbitrary, and line networks in Fig. \ref{fig:networks} by ensemble averaging empirical cost of the ODOL algorithm over $100$ independent trials. Note that if the associated network is a tree network, e.g., the star network is a tree, the same cost could be attained through the OEDOL algorithm as shown analytically in Section 3. We point out that Fig. \ref{fig:comp_team} is not a time evolution of a learning algorithm. In particular, for each time horizon $T$, we compare the corresponding oracle cost over certain networks. As pointed out before, because of the most possible connection between the agents, the fully connected network results in the least oracle cost for each time horizon $T$ while the line network possesses the largest cost. Additionally, because of the cumulation of the impact of the actions at each time, for larger time horizon the corresponding oracle cost is also larger, i.e., $J(T)$ is a non-decreasing function of $T$ (indeed increasing function since measurement noise is a white process). The arbitrary network is relatively more connected. In particular, in the arbitrary network, neighborhood size of agents is chosen uniformly between 2 and 10 while in the star network, neighborhood size of agents except agent-$1$ is $2$. We observe that oracle cost over the arbitrary network is smaller than the oracle cost over the star network. For $T=1$, all network configurations lead to the same oracle cost since agents can exchange information beginning at time $t=2$. Furthermore, in Fig. \ref{fig:comp_team}, for time horizon $T=2$, the oracle cost in the star network is larger than the cost in the line network. This is because even though in the star network any two agents are connected through at most $2$ hops, over a horizon $T=2$, the information can only propagate over $1$ hop and in the line network all the agents except agents $1$ and $2$ have 2 neighbors each. For horizon $T>2$, the information can now propagate over more than $1$ hop and correspondingly, the cost in the star network is less than the cost in the line network.

In addition to the comparison of oracle cost, $J(T)$, for different time horizons over certain networks, in Fig. \ref{fig:comp_terminal}, we compare the terminal cost, $P(T)$, in order to observe the cumulative impact of the actions in the beginning stages of the horizon, e.g., $t \ll T$. We again ensemble averaged empirical cost of the ODOL algorithm over 100 independent trials. Even though a line and a fully connected network are two extreme connections for a network, in Fig. \ref{fig:comp_terminal}, the terminal costs, $P(T)$, are asymptotically, as $T \rightarrow \infty$, the same. We recall that this is pointed out in Remark 3.1 and attributed to the negligible difference between the available information at each agent over fully connected and line networks as the information sets grow. We also point out that in the zoomed-out plot, for $T=2$, the terminal cost in the star network is larger than the terminal cost in the line network while the terminal cost in star network becomes substantially smaller for larger $T$ as explained above for Fig. \ref{fig:comp_team}.

Importantly, in Figs. \ref{fig:comp_team} and \ref{fig:comp_terminal}, the oracle and terminal costs lead to counter-intuitive results as time horizon grows. In particular, while the oracle costs over different networks are separated as time horizon grows, the terminal costs converge to each other. The cumulation of the costs of each action plays a significant role in these distinct results. By yielding relatively large cost, the actions in the early stages have crucial impact on the team cost. Therefore, the sequential algorithms aiming to learn the underlying state in time through a trial and error based approach lead to larger team cost over horizon since trials at early stages lead to significantly large costs. In the following, we analyze such circumstances.

\begin{figure}[t!]
  \centering
  \subfloat[Arbitrary network]{\includegraphics[width=0.5\textwidth]{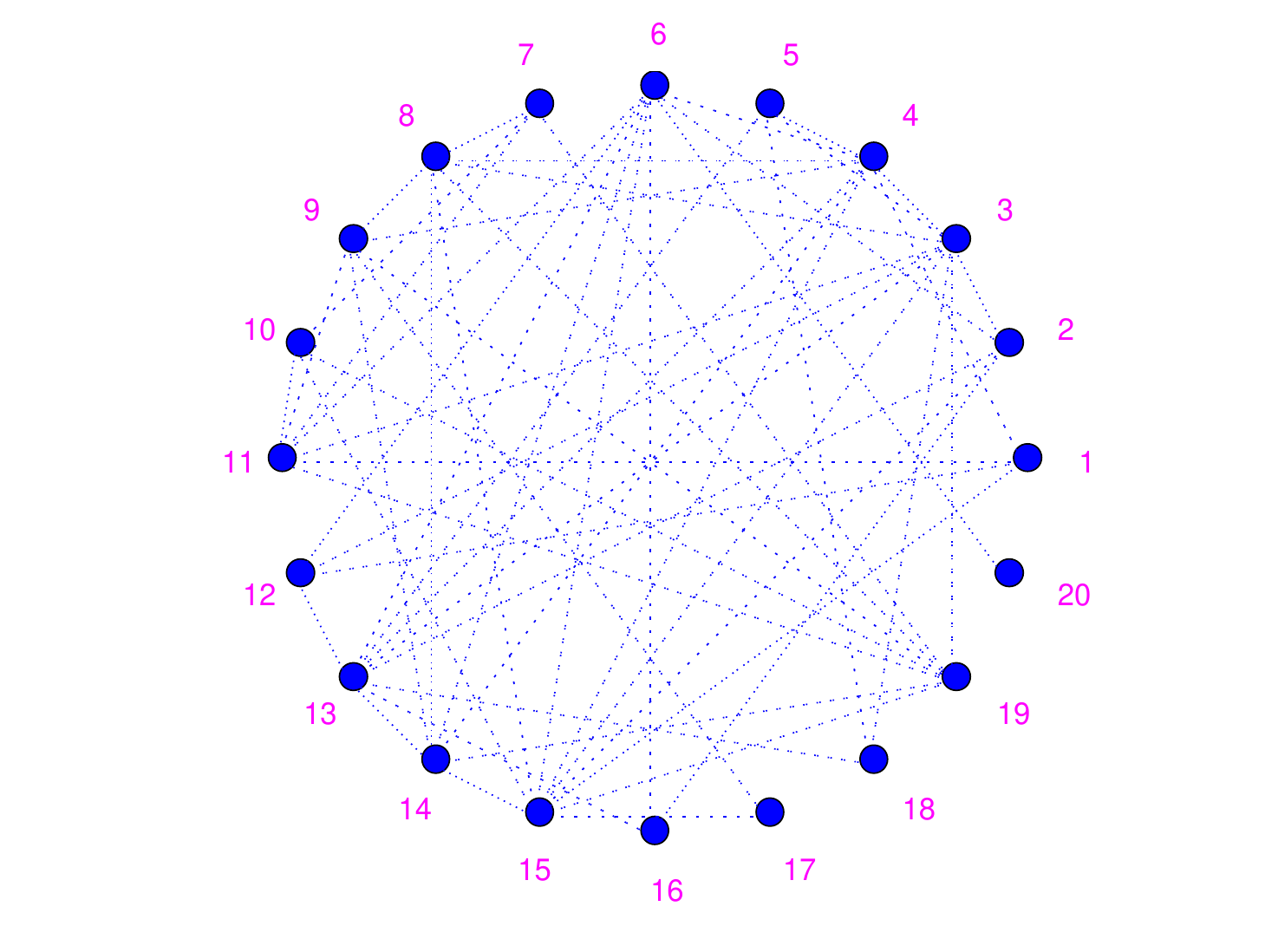}\label{fig:a_2}}
  \hfil
  \subfloat[Corresponding spanning tree]{\includegraphics[width=0.5\textwidth]{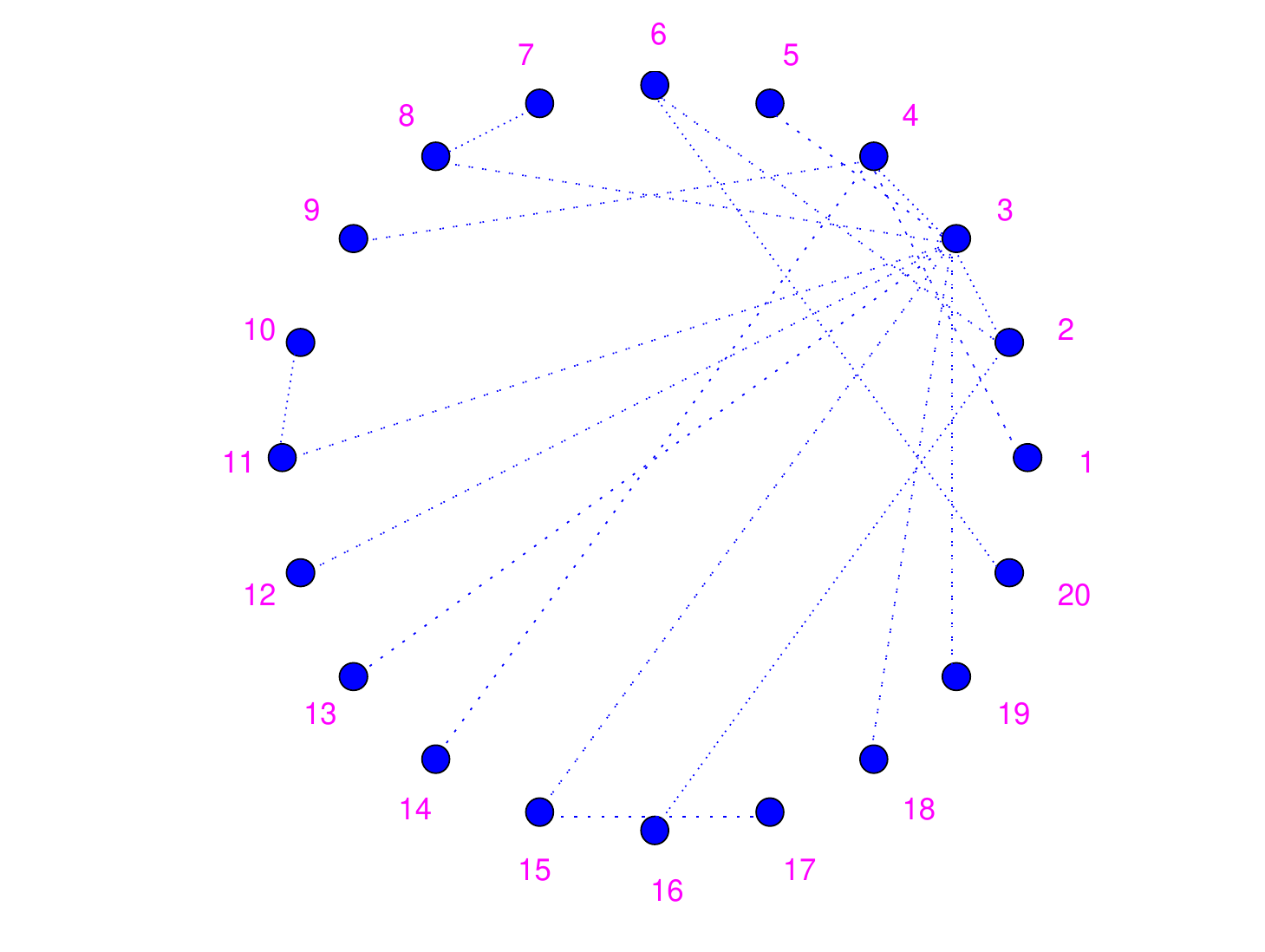}\label{fig:s_2}}
  \caption{An arbitrary network and a corresponding spanning tree.}
  \label{fig:networks2}
\end{figure}

\begin{figure}
  \centering
  \includegraphics[width=0.7\textwidth]{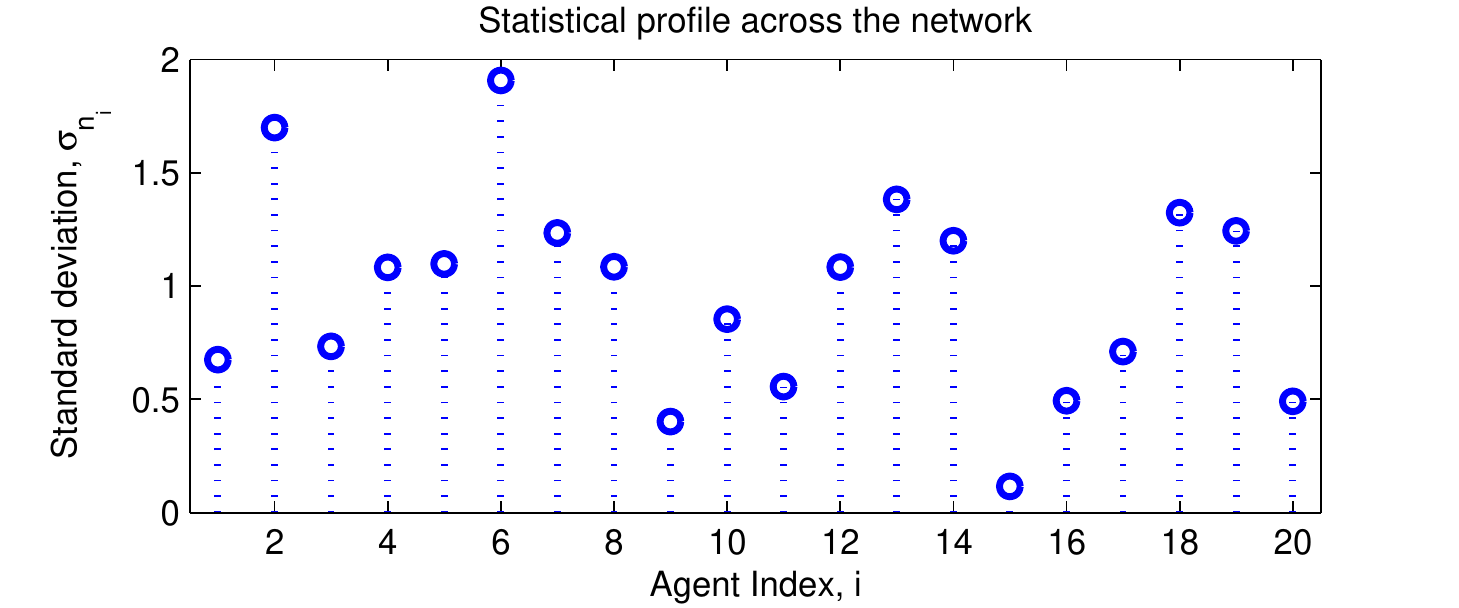}\\
  \caption{Standard deviation of measurement noise across the network in Fig. \ref{fig:networks2}.}\label{fig:noises2}
\end{figure}

We consider a randomly generated network in Fig. \ref{fig:a_2}, constructed as the network in Fig. \ref{fig:n_a}. This network is not a tree and has loops. We construct a spanning tree of this network, see Fig. \ref{fig:s_2}, based on the paths from the agent having the largest neighborhood size, i.e., $i = \arg \max_j \pi_j$, and then eliminating the multi-paths \cite{wu}. The statistical profile of the measurement noise is plotted in Fig. \ref{fig:noises2}. Here, we consider the diffusion recursive least squares (D-RLS) algorithm \cite{cattivelli2008} with Laplacian combination rule \cite{saber2004} for the incremental update and relative-variance combination rule for the spatial update. In D-RLS, agents use a recursive least squares (RLS) update locally, diffuse local estimate to the neighbors at each time instant, and combine received estimates and the local one linearly through certain combination weights, e.g., agent-$i$ has the weight $\lambda_{i,j}$ for the information received from the neighbor-$j$:
\begin{equation}\label{eq:relative}
\lambda_{i,j} = \left\{\begin{array}{ll} \frac{\sigma_{n_j}^{-2}}{\sum_{k\in\cN_i \cup \{i\}}\sigma_{n_k}^{-2}} & \mbox{if $j \in \cN_i\cup\{i\}$}\\ 0 & \mbox{otherwise}.\end{array}\right.
\end{equation}

\begin{figure}[t!]
  \centering
  \includegraphics[width=.9\textwidth]{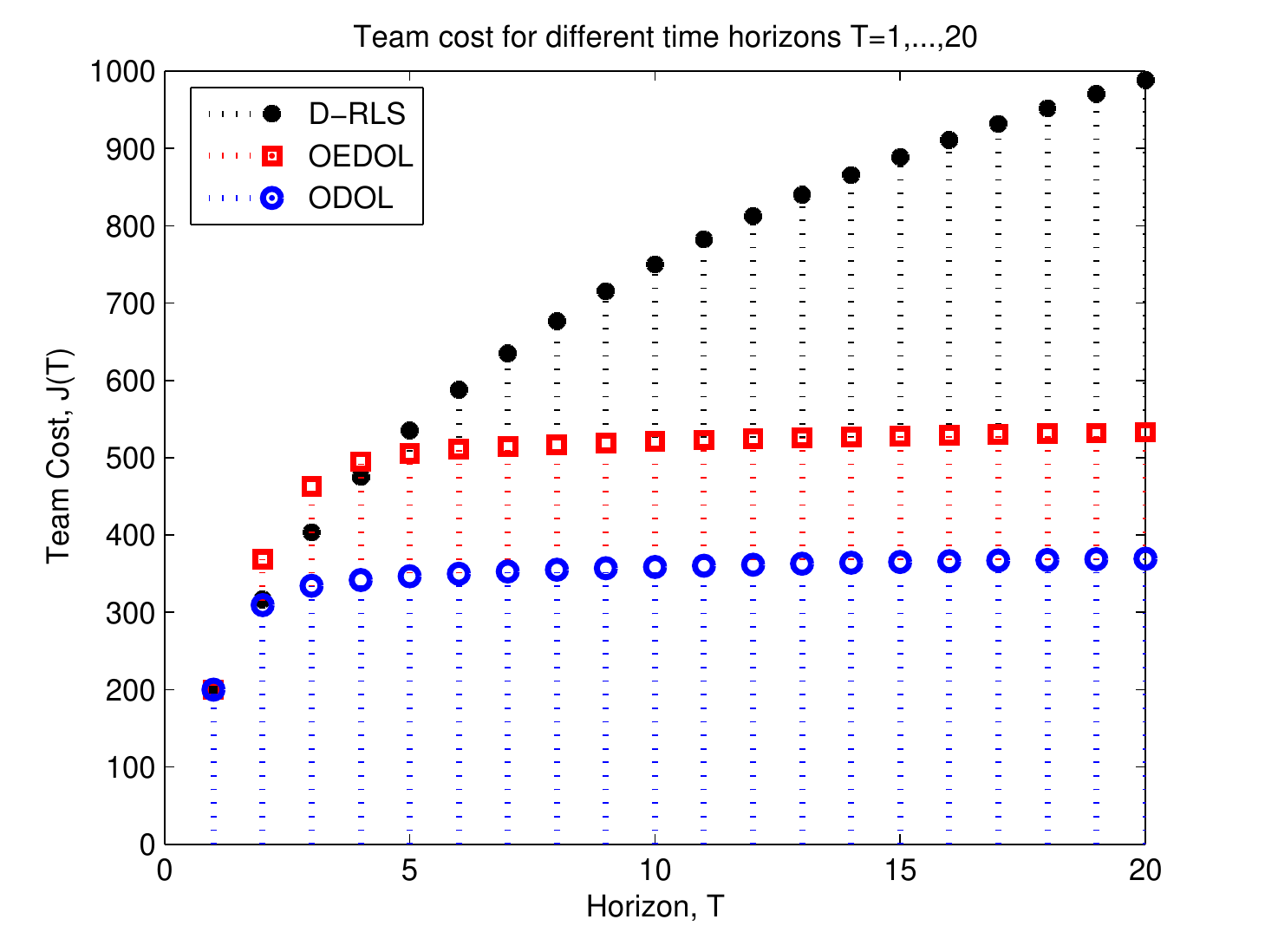}\\
  \caption{Comparison of team cost of the algorithms for different time horizon.}\label{fig:oedol_team}
\end{figure}

\begin{figure}[t!]
  \centering
  \includegraphics[width=.9\textwidth]{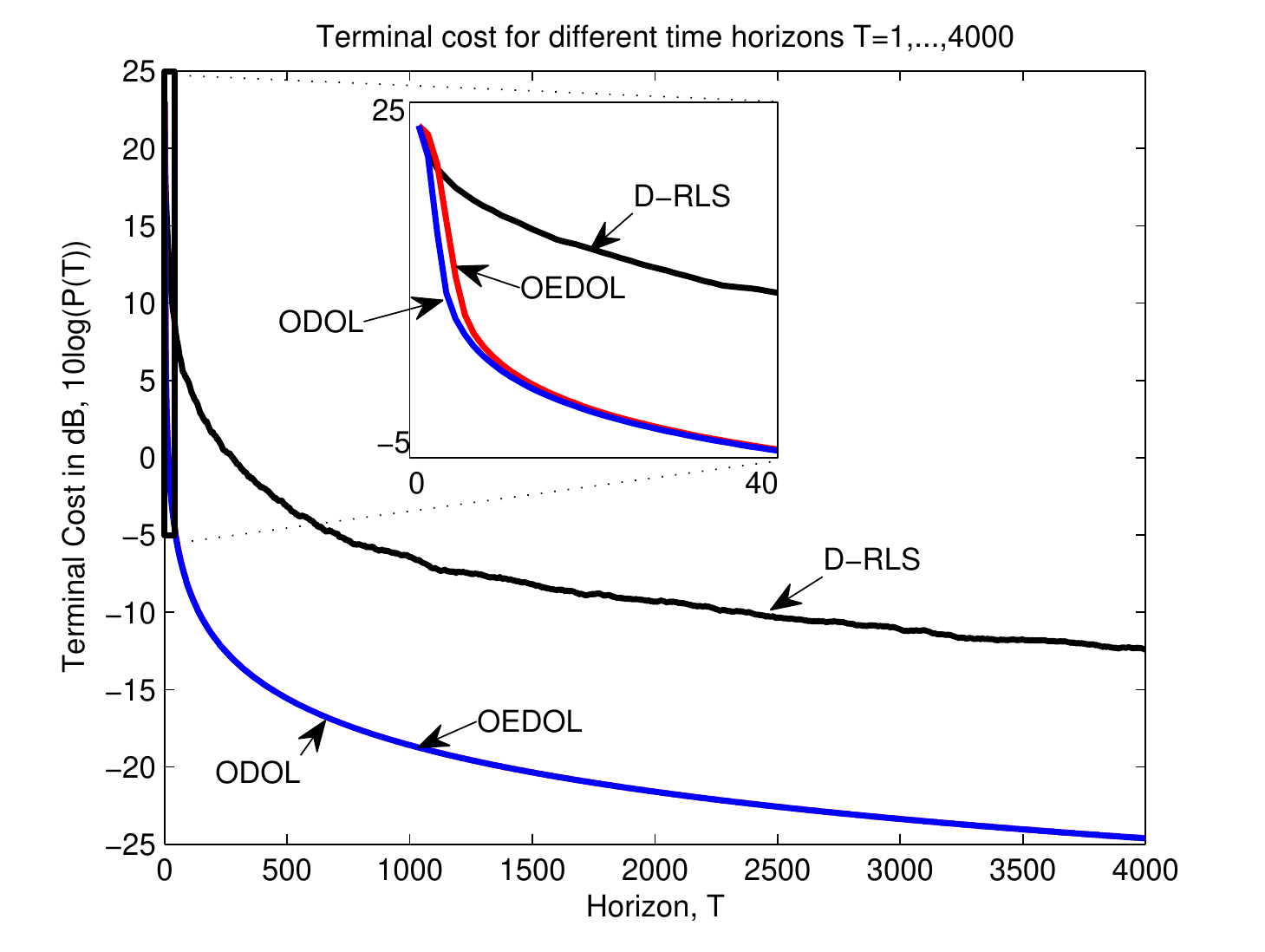}\\
  \caption{Comparison of terminal cost of the algorithms for different time horizon.}\label{fig:oedol_terminal}
\end{figure}

In Fig. \ref{fig:oedol_team}, we compare the ensemble average of the empirical-cost (over 100 independent trials) of the ODOL and D-RLS algorithms over Fig. \ref{fig:a_2} and the OEDOL algorithm over Fig. \ref{fig:s_2} for different time horizon. Note that in the OEDOL algorithm, agents exchange local estimates while in the ODOL algorithm, agents exchange information through time-stamped approach. The OEDOL algorithm can achieve the oracle performance over a tree network. Therefore, we construct a spanning tree of the arbitrarily generated network and the OEDOL algorithm achieves the oracle performance of the spanning tree. Since we eliminate certain links while constructing the spanning tree, the oracle performance, i.e., team-cost, is worse than the oracle performance over the original network. However, even though the OEDOL algorithm operates over the spanning tree, the algorithm has performed better than the D-RLS algorithm operating over the original network. Additionally, in Fig. \ref{fig:oedol_terminal}, we compare the terminal cost of the algorithms for larger time horizon, e.g., $T=1,\cdots,4000$. We observe that even though the OEDOL algorithm operates over the spanning tree, the terminal costs of the ODOL and OEDOL algorithms are close to each other for large time horizon, e.g., $T > 40$. We also note that as $T\rightarrow \infty$, the terminal cost decreases since noise processes are white.

\begin{figure}[t!]
  \centering
  \begin{overpic}[width=0.9\textwidth]{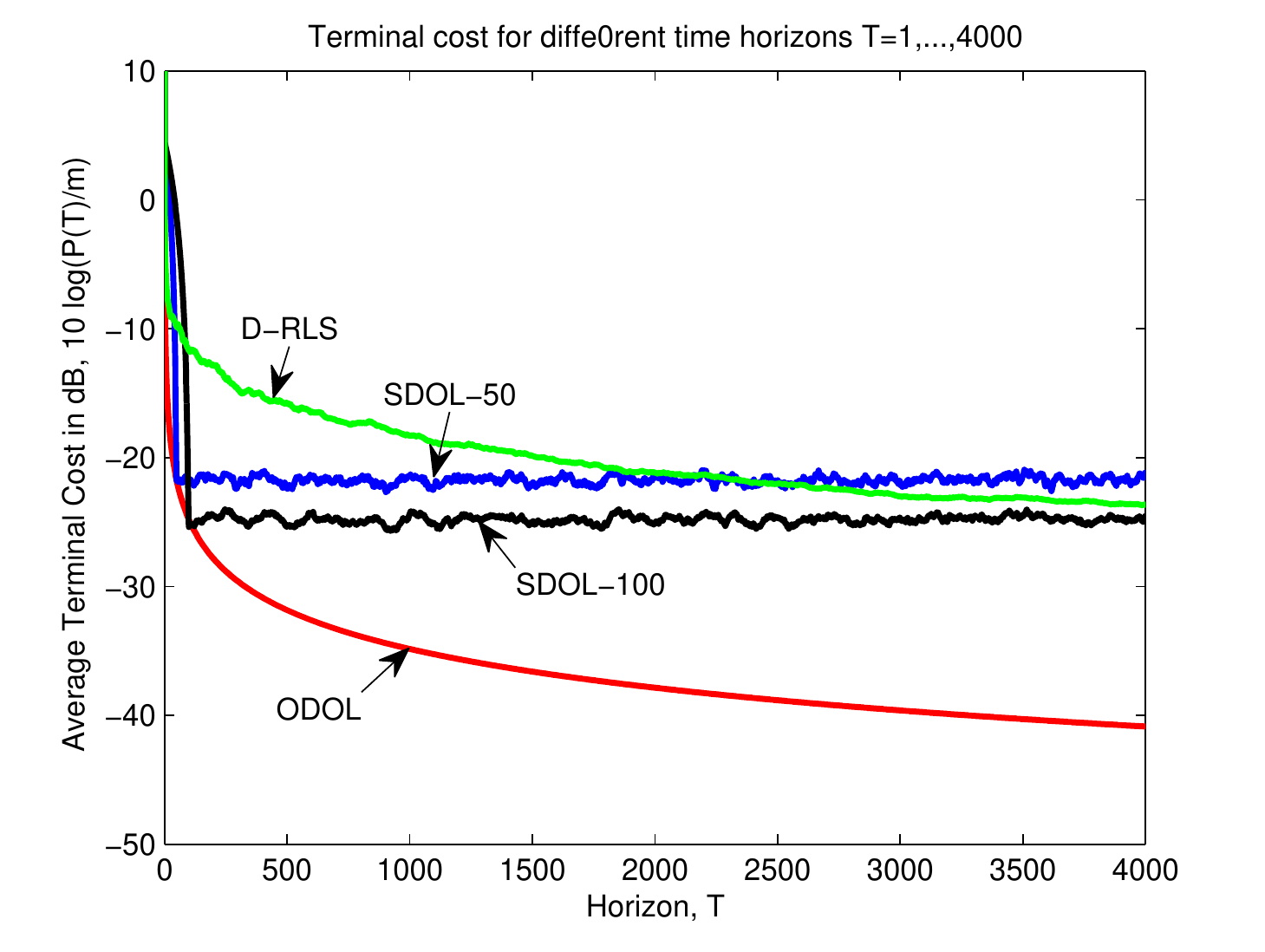}
     \put(55,40){\includegraphics[scale=0.25]{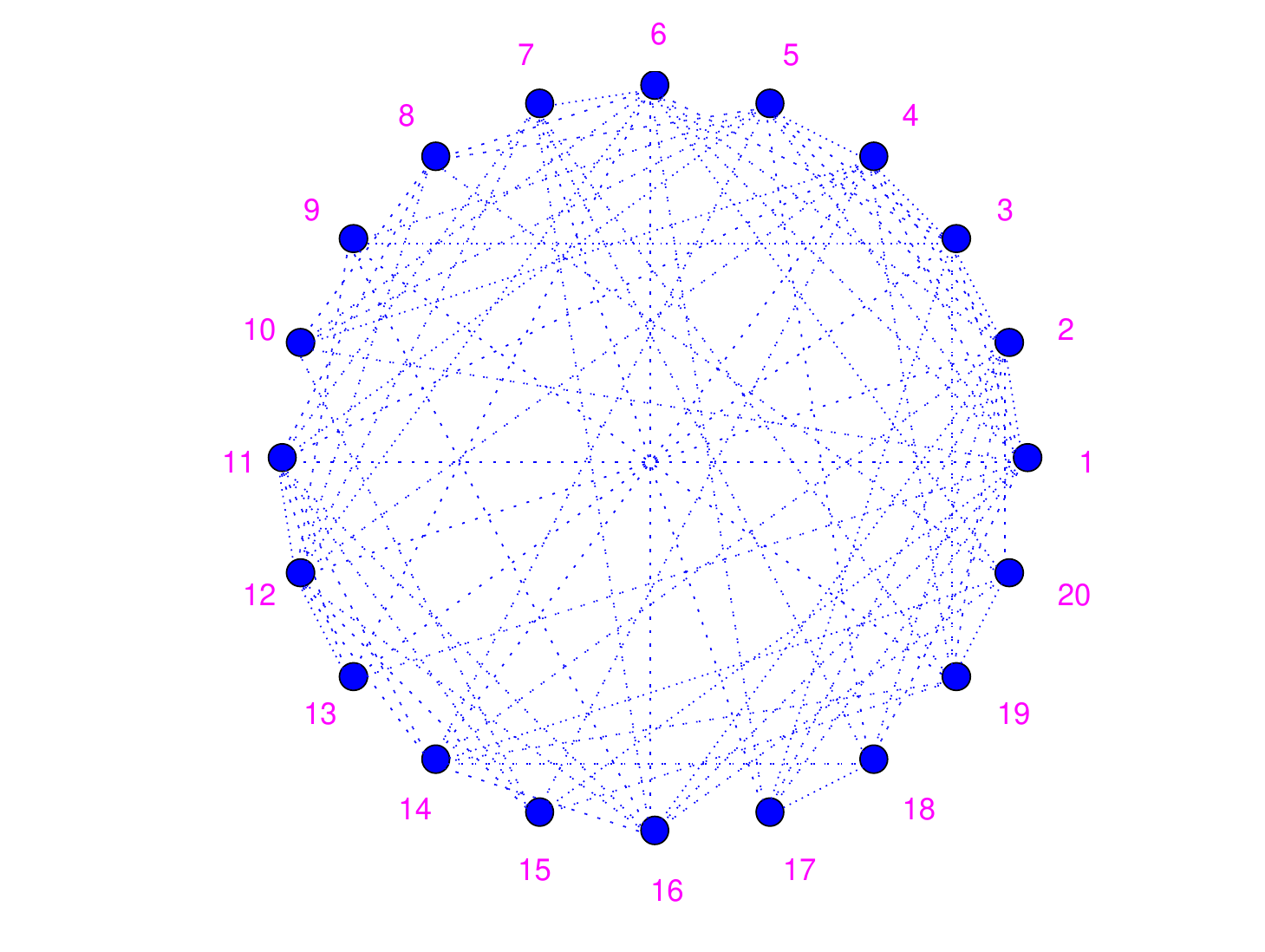}}
  \end{overpic}\\
  \caption{Comparison of terminal cost of the SDOL-50 and SDOL-100 algorithms.}\label{fig:sdol}
\end{figure}

Finally, in Fig. \ref{fig:sdol} we examine the impact of the time-windowing depth, e.g., $T_o=50,100$, on the terminal cost for different time horizon over an arbitrarily generated network of $20$ agents seen on the upper-right of Fig. \ref{fig:sdol}. We note out that for $T< T_o$, the terminal cost of the SDOL algorithms differs from the terminal cost of the ODOL algorithm, i.e., the oracle performance since in the SDOL algorithms, agents uses the update \eqref{eq:update} for $t<T_o$ assuming that all measurements $y_{j,\tau} = 0$ for $j=1,\cdots,m$ and $\tau < 1$. As also pointed out in Remark 4.1, by excluding the impact of these measurements on the future actions through sliding time-window, the terminal cost of the SDOL-$T_o$ algorithm eventually reaches the terminal cost of the ODOL algorithm at $T=T_o$ and for larger time window, this performance is maintained. Furthermore, depending on the time window depth $T_o$, the agents can achieve smaller terminal cost.

\section{Conclusion}
Distributed algorithms have attracted significant attention due to their wide spread applicability to highly complex structures from biological systems to social and economical networks. However, there are still challenges for disclosure and utilization of information among agents. We have considered this problem as a team problem, in which each agent takes actions, e.g., which information to disclose and how to construct the local estimate. We introduced the ODOL algorithm achieving the oracle cost for Gaussian state and noise statistics through a time-stamped approach. Importantly, we have shown that the exchange of the local estimates is sufficient to achieve the oracle cost only over certain network topologies, e.g., over the introduced tree networks involving cell structures. Furthermore, we have introduced the OEDOL algorithm, which achieves the oracle cost through the exchange of local estimates over tree networks. Finally, we have introduced a time-windowing approach for practical applications due to reduced complexity. The sub-optimal approaches possess recursive updates with time-invariant combination weights that should be calculated only once.

Some future directions of research on this topic include team problems for distributed MMSE estimation of a dynamic state and formulation of optimal information exchange in team decision problems for linear continuous-time systems \cite{basar80}.

\section{Acknowledgments}
 This work is supported in part by TUBITAK Contract No 115E917, in part by the U.S. Office of Naval Research (ONR) MURI grant N00014-16-1-2710, and in part by NSF under grant CCF 11-11342.


\section*{References}

\bibliography{my_references}

\end{document}